\pdfoutput=1

\documentclass[useAMS,usenatbib,fleqn]{mnras}

\usepackage{graphicx}
\usepackage{amsmath,amssymb}
\usepackage{natbib}

\title[Orbit classification in a non-spinning binary black hole system]{Orbit classification in an equal-mass non-spinning binary black hole pseudo-Newtonian system}

\author[E.E. Zotos et al.]{Euaggelos E. Zotos$^1$\thanks{E-mail: evzotos@physics.auth.gr}, F.L. Dubeibe$^2$\thanks{E-mail: fldubeibem@unal.edu.co} and Guillermo A. Gonz\'{a}lez$^3$\thanks{E-mail: guillermo.gonzalez@saber.uis.edu.co} \\
$^1$ Department of Physics, School of Science, Aristotle University of Thessaloniki, GR-541 24, Thessaloniki, Greece \\
$^2$ Grupo de Investigaci\'{o}n Cavendish, Facultad de Ciencias Humanas y de la Educaci\'{o}n, Universidad de los Llanos, \\ Villavicencio 500017, Colombia \\
$^3$ Grupo de Investigaci\'{o}n en Relatividad y Gravitaci\'{o}n, Escuela de F\'{i}sica, Universidad Industrial de Santander, \\ A.A. 678, Bucaramanga 680002, Colombia.
}

\begin{document}

\date{Accepted 2018 April 12. Received 2018 April 9; in original form 2018 February 2}

\pubyear{2018} \volume{477} \pagerange{5388--5405}

\setcounter{page}{5388}

\maketitle

\label{firstpage}

\begin{abstract}
The dynamics of a test particle in a non-spinning binary black hole system of equal masses is numerically investigated. The binary system is modeled in the context of the pseudo-Newtonian circular restricted three-body problem, such that the primaries are separated by a fixed distance and move in a circular orbit around each other. In particular, the Paczy\'{n}ski-Wiita potential is used for describing the gravitational field of the two non-Newtonian primaries. The orbital properties of the test particle are determined through the classification of the initial conditions of the orbits, using several values of the Jacobi constant, in the Hill's regions of possible motion. The initial conditions are classified into three main categories: (i) bounded, (ii) escaping and (iii) displaying close encounters. Using the smaller alignment index (SALI) chaos indicator, we further classify bounded orbits into regular, sticky or chaotic. To gain a complete view of the dynamics of the system, we define grids of initial conditions on different types of two-dimensional planes. The orbital structure of the configuration plane, along with the corresponding distributions of the escape and collision/close encounter times, allow us to observe the transition from the classical Newtonian to the pseudo-Newtonian regime. Our numerical results reveal a strong dependence of the properties of the considered basins with the Jacobi constant as well as with the Schwarzschild radius of the black holes.
\end{abstract}


\section{Introduction}
\label{intro}

The planar circular restricted three-body problem (henceforth PCRTBP) was first formulated in the realm of Newtonian dynamics as a very particular case of the general three-body problem \citep{E67}. It has been widely investigated in the literature and it can serve as an example of classical chaos \citep{H03}. After a series of papers published by Str\"{o}mgren and colleagues in the Copenhagen Observatory, where the primaries of the PCRTBP were chosen to have equal masses \citep[see e.g.,][]{BS15}, the problem is nowadays known as the Copenhagen problem. Despite its simplicity and the extensive studies carried out in the Copenhagen Observatory, the Copenhagen problem still triggers numerical explorations due to its astrophysical relevance. Just to name a few works along this line, we mention the studies related to its non-integrability in Poincar\'{e}'s sense \citep[e.g.,][]{S12}, the numerical evidence about the structure of scattering functions and their connection with primitive periodic orbits \citep[e.g.,][]{BTS96}, or the classification of orbits performed in \citet{N04,N05} and \citet{dAT14}.

In the extreme cases of compact objects (e.g. black holes, white dwarfs or neutron stars), where the Newtonian mechanics is no longer valid, a fully general relativistic treatment or relativistic corrections to the PCRTBP are needed. Several efforts have been made in this direction, but to our knowledge, the Copenhagen problem has never been considered in this context. Some related papers about the first-order post-Newtonian approximation of the PCRTBP, include the studies about the effects on the dynamics of the distance between the primaries conducted in \citet{HW14}, the demonstration of the equivalence between the Lagrangian and Hamiltonian approaches and the conservation of the Jacobi integral in the post-Newtonian PCRTBP \citep{DLCG17a}, and the studies about the orbit classification in the post-Newtonian planar circular restricted Sun-Jupiter system \citep{ZD18}.

On the other hand, to avoid the cumbersome equations of motion taking place in the post-Newtonian formalism and to account for additional relativistic effects, like the presence of event horizons, we can use pseudo-Newtonian approximations. For example, in \citet{SL06} they used the Paczy\'{n}ski-Wiita potential, to study the dynamics of the PCRTBP in the Hill's approximation. Moreover, in a recent paper \citep{DLCG17b}, we used the Fodor-Hoenselaers-Perj\'{e}s procedure \citep{FHP89} to perform an expansion in the mass potential with the aim to study the dynamics of the PCRTBP, in the context of a pseudo-Newtonian approximation. Following this idea, in the present paper, we study how the existence of an event horizon affects the dynamics of the Copenhagen problem. To do so, we mimic the relativistic effects of the event horizon by modeling the gravitational field of the two primaries (black holes) in the Copenhagen problem through the Paczy\'{n}ski-Wiita potential.

In the present work, we choose the Paczy\'{n}ski-Wiita potential to model the Copenhagen problem for three main reasons: (i) its simplicity and practicality; (ii) it is a standard tool in relativistic astrophysics; and (iii) it reproduces some features of the most important processes taking place close to Schwarzschild black holes, e.g., the location of the innermost stable circular orbit, the location of the marginally bound orbit, and the form of the Keplerian angular momentum \citep{A09}. Probably the most glaring shortcoming of the Paczy\'{n}ski-Wiita potential is that it was specifically designed for approximately reproducing relativistic effects of circular orbits around single black holes. For this reason, some other pseudo-Newtonian potentials for static spherically symmetric black hole metrics has been recently proposed \citep[see e.g.,][]{TR13,TR14,W12,WL17}. These new potentials are able to reproduce several relativistic features with higher accuracy than the Paczy\'{n}ski-Wiita potential, nevertheless, the gain in accuracy brings as a consequence more elaborate potentials that include velocity-dependent terms, which significantly complicates the implementation of the two-body problem. Therefore, taking into account that our idea is to perform a preliminary systematic study of the orbital behavior around a non-Newtonian binary system, we shall use the Paczy\'{n}ski-Wiita potential instead of any other pseudo-Newtonian potential.

The paper is organized as follows: In Section \ref{des} we describe the main properties of the dynamical system. In Section \ref{eqps} we present a heuristic analysis of the equilibrium points and the Hill's regions configurations. In the following section, we outline the computational methods used for the classification of the initial conditions of the orbits. Section \ref{orbclas}, contains all the numerical results on the pseudo-Newtonian binary system of two black holes. Finally, the main conclusions of our numerical study are emphasized in Section \ref{conc}.

\section{Description of the astrophysical system}
\label{des}

\begin{figure}
\centering
\resizebox{\hsize}{!}{\includegraphics{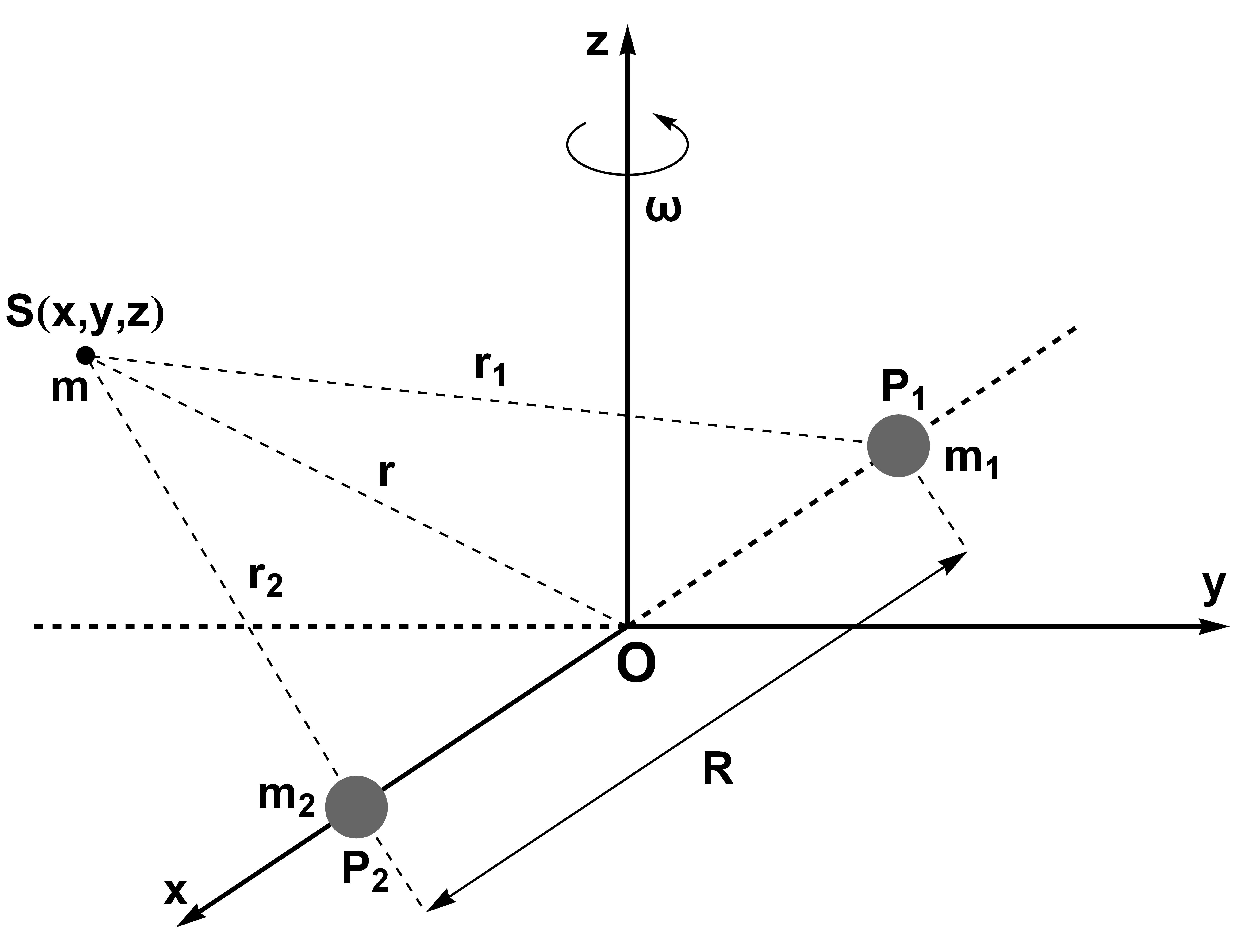}}
\caption{The configuration of two primary bodies, $P_1$ and $P_2$, moving in circular orbits about their common center of gravity (origin), unperturbed by the test particle of infinitesimal mass, in the restricted three-body problem.}
\label{sch}
\end{figure}

According to the theory of the classical circular restricted three-body problem (CRTBP), the two primary bodies $P_1$ and $P_2$ move on circular orbits, with the same angular velocity $\omega$, about their center of mass (see Fig. \ref{sch}). The third body moves in coplanar orbits, under the action of the gravitational field of the primaries and does not affect their motion.

The condition for the angular velocity to keep the primaries moving in circular Keplerian orbits at the equatorial plane reads as
\begin{equation}
\omega = \sqrt{\frac{G\left(m_1 + m_2\right)}{R^3}},
\label{ang}
\end{equation}
where $R = x_1 + x_2$, denotes the distance between the primaries.

As noted in the Introduction, the Paczy\'{n}ski-Wiita potential (see e.g., \citet{PW80} and \citet{A09}) is a very practical pseudo-Newtonian potential that may accurately model general relativistic effects, related with the motion of matter near a non-spinning black hole. Unlike most pseudo-Newtonian potentials, this is not an ad hoc potential that can be derived by a step-by-step formal procedure (as shown by \cite{A09}), and allow us to calculate all quantities of interest in a simple closed form. Such potential modifies the typical $G M/r$ Newtonian potential by $GM/(r - r_S)$, where $r_S$ is the relativistic Schwarzschild radius. Thus, the pseudo-Newtonian effective potential that describes a non-spinning binary black hole system can be written as
\begin{equation}
U(x,y,z) = \frac{G m_1}{r_1 - r_{S1}} + \frac{G m_2}{r_2 - r_{S2}} + \frac{\omega^2}{2}\left(x^2 + y^2\right),
\label{eff}
\end{equation}
where
\begin{align}
r_1 &= \sqrt{\left(x - x_1\right)^2 + y^2 + z^2}, \nonumber\\
r_2 &= \sqrt{\left(x - x_2\right)^2 + y^2 + z^2},
\label{dist}
\end{align}
are the distances to the respective primaries, while
\begin{equation}
r_{S1} = \frac{2G m_1}{c^2}, \ \ \ r_{S2} = \frac{2G m_2}{c^2},
\label{rad}
\end{equation}
are the Schwarzschild radii for the primaries $P_1$ and $P_2$, respectively, where $c$ is the speed of light. In order to avoid that the horizons of the massive bodies do intersect and to exclude velocities above $c$, the value of the Schwarzschild radius is set to the interval $[0, 0.38]$. It should be noted that if we set $r_S = 0$ to Eq. (\ref{eff}) the effective potential is automatically reduced to the classical Newtonian problem.

In the case of pseudo-Newtonian dynamics the conditions for circular Keplerian orbits, of the Copenhagen problem, reads
\begin{equation}
\omega = \frac{R}{R - r_S}\sqrt{\frac{2GM}{R^3}},
\label{om1}
\end{equation}
with $M = m_1 = m_2$. In Appendix \ref{appex} we prove the existence of circular orbits in the case where the two primaries are modelled by a Paczy\'{n}ski-Wiita potential. Along the paper we shall use the Szebehely's notation to nondimensionalize the problem, i.e. we introduce the mass parameter $\mu$, such that $G m_1 = 1 - \mu$ and $G m_2 = \mu$. For the description of the dynamical system we choose, as a frame of reference, a synodic barycenter rotating coordinate system where its origin $O$ is at the center of mass of the primaries, while the centers of the primaries are located at $(x_1, 0)$ and $(x_2,0)$, where $x_1 = - \mu$ and $x_2 = 1 - \mu$. Since we are interested in the Copenhagen problem, in all that follows we set $\mu = 1/2$.

In terms of the parameters used along this paper, the angular velocity (\ref{om1}) takes the form
\begin{equation}
\omega = \frac{1}{1 - r_S}.
\label{om2}
\end{equation}
Therefore, the whole information about the binary system and hence of $r_S$, relies on the value of $c$. In other words, from the relation $v = \omega R$ for circular orbits, and by dividing both sides by $c$, we have $v/c = \omega/c$, due to the fact that $R = x_1 + x_2 = 1 - \mu + \mu = 1$, thus
\begin{equation}
\frac{v}{c} = \frac{\omega}{c} = \frac{c}{c^2 - 1}.
\label{v0}
\end{equation}
So, for example, for a binary system with orbital velocities of the order $v/c \approx 10^{-4}$ (note that $v/c$ is a non-dimensional quantity), we replace this value in the left-hand side of Eq. (\ref{v0}) and solving for $c$, we obtain $c \approx 10^{4}$. Now, substituting this value in the Schwarzschild radius we get $r_S \approx 10^{-8}$.

At this point, it should be emphasized that according to Einstein's theory of General Relativity, accelerating masses cause gravitational waves (GWs) \citep[see e.g.,][]{L92}. In particular, a gravitational system composed of two compact objects orbiting each other in a bound binary system is a prominent source of GWs, as the constituents accelerate in their orbits \citep{H02,S13}. The emission of GW would cause the loss of energy in the system and consequently, the compact objects would eventually approach each other. Therefore, in a realistic context of binary systems, the GW emission in our system could become relevant for the orbital dynamics. If this is the case, the ratio between the Schwarzschild radius and the separation of the primaries should satisfy $r_S/R < 10^{-4}$, in order to consider negligible the GW emission for simulations of the order of thousand orbital times. In spite of this limitation, yet aiming to observe significant differences in the orbital properties of the test particle, we will use values of $r_S$ up to $10^{-2}$ in our numerical calculations.

In the synodic frame of reference the motion of a test particle can be described by the following set of differential equations
\begin{align}
\ddot{x} - 2 \omega \dot{y} = \frac{\partial U(x,y,z)}{\partial x}, \nonumber\\
\ddot{y} + 2 \omega \dot{x} = \frac{\partial U(x,y,z)}{\partial y}, \nonumber\\
\ddot{z} = \frac{\partial U(x,y,z)}{\partial z}.
\label{eqmot}
\end{align}

In the same vein, the variational equations (needed for the computation of the SALI) are given by
\begin{align}
\dot{(\delta x)} &= \delta \dot{x}, \nonumber\\
\dot{(\delta y)} &= \delta \dot{y}, \nonumber\\
\dot{(\delta z)} &= \delta \dot{z}, \nonumber\\
(\dot{\delta \dot{x}}) &= \frac{\partial^2 U}{\partial x^2}\delta x + \frac{\partial^2 U}{\partial x \partial
y}\delta y + \frac{\partial^2 U}{\partial x \partial z}\delta z + 2 \omega \delta \dot{y}, \nonumber \\
(\dot{\delta \dot{y}}) &= \frac{\partial^2 U}{\partial y \partial x}\delta x + \frac{\partial^2 U}{\partial
y^2}\delta y + \frac{\partial^2 U}{\partial y \partial z}\delta z - 2 \omega \delta \dot{x}, \nonumber\\
(\dot{\delta \dot{z}}) &= \frac{\partial^2 U}{\partial z \partial x}\delta x + \frac{\partial^2 U}{\partial z
\partial y}\delta y + \frac{\partial^2 U}{\partial z^2}\delta z.
\label{variac}
\end{align}

The set of the equations of motion (\ref{eqmot}) admits only the following integral of motion (also known as the Jacobi integral)
\begin{equation}
J(x,y,z,\dot{x},\dot{y},\dot{z}) = 2U(x,y,z) - \left(\dot{x}^2 + \dot{y}^2 + \dot{z}^2\right) = C,
\label{ham}
\end{equation}
where $\dot{x}$, $\dot{y}$, and $\dot{z}$ are the velocities conjugate to $x$, $y$, and $z$, respectively, while $C$ is the numerical value of the Jacobi constant which is conserved.

\begin{figure}
\centering
\resizebox{\hsize}{!}{\includegraphics{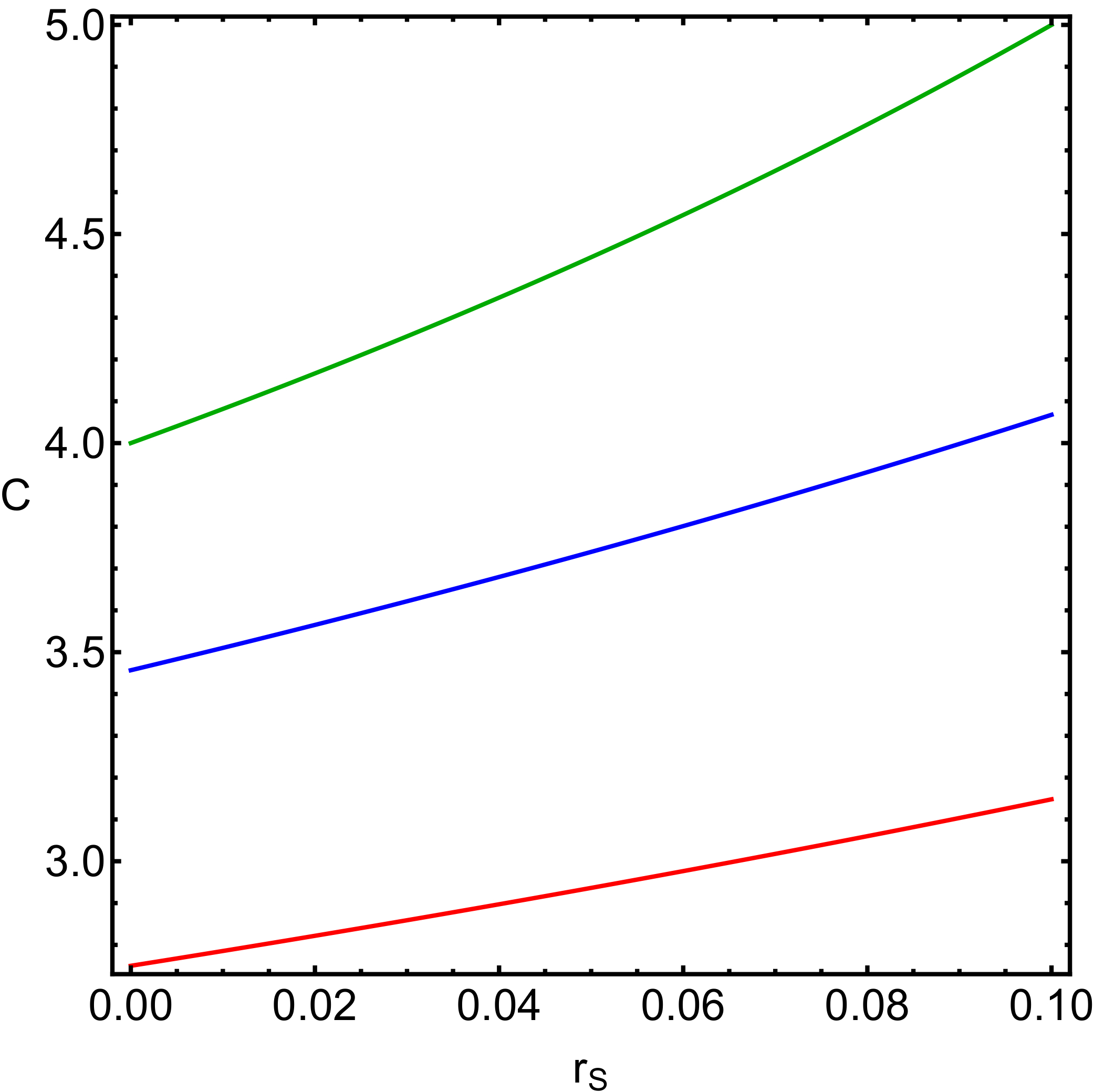}}
\caption{Parametric evolution of the critical values of the Jacobi constant, $C_1$ (green), $C_2 = C_3$ (blue), and $C_4 = C_5$ (red), as a function of the Schwarzschild radius, when $r_S \in [0, 0.1]$. (Color figure online.)}
\label{crits}
\end{figure}

\section{Equilibrium points and Hill's regions configurations}
\label{eqps}

\begin{figure*}
\centering
\resizebox{\hsize}{!}{\includegraphics{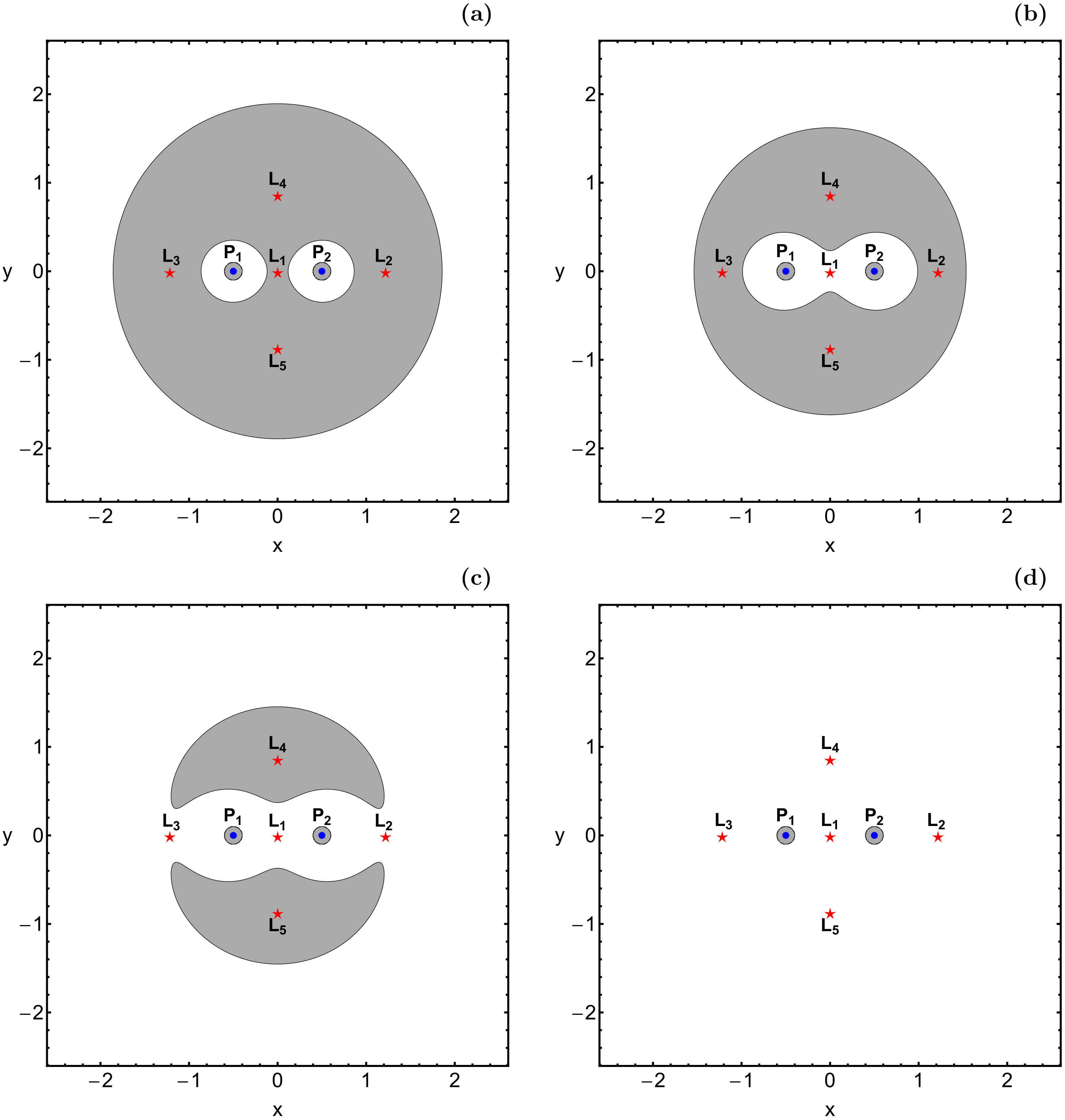}}
\caption{The structure of the Hill's regions configurations, when $r_S = 0.1$. The white zones correspond to the allowed regions of motion, gray zones indicate the forbidden regions, while the boundary between the gray and the white regions denotes the corresponding ZVC. The blue points indicate the position of the primaries $P_1$ and $P_2$, while the red stars point out the location of the Lagrange points. The value of the Jacobi constant is set as: (a): $C = 5.5$; (b): $C = 4.5$; (c): $C = 4$; (d): $C = 3$. (Color figure online.)}
\label{conts}
\end{figure*}

In the binary system of two black holes, described by the effective potential (\ref{eff}), there exist five equilibrium points at which
\begin{equation}
\frac{\partial U}{\partial x} = \frac{\partial U}{\partial y} = \frac{\partial U}{\partial z} = 0.
\label{lps}
\end{equation}
All five equilibria (also known as Lagrange points) lie on the configuration $(x,y)$ plane with $z = 0$. The points $L_1$, $L_2$, and $L_3$ are located on the $x$-axis and they are called collinear points. On the other hand, the points $L_4$ and $L_5$ are located on the vertices of two equilateral triangles and they are called triangular points.

The values of the Jacobi constant at the equilibrium points are in fact critical energy levels. Obviously, the critical values of the energy are functions of the Schwarzschild radius $r_S$. In Fig. \ref{crits} we see how the critical values $C_1$ (green), $C_2 = C_3$ (blue), and $C_4 = C_5$ (red) evolve, when $r_S \in [0, 0.1]$.

For the case of the planar problem, where $z = \dot{z} = 0$, the Hill's regions of allowed motion are the subsets of the configuration space, resulting from the projection of the four-dimensional energy manifold, onto the $(x,y)$ plane. Each Hill's region configuration is divided into three domains: (i) the interior region, where $x(L_3) < x < x(L_2)$, the exterior region, where $x < x(L_3)$ or $x > x(L_2)$, and (iii) the energetically forbidden regions, where $2U(x,y,z) < C$. Furthermore, the Hill's regions are bounded by the so-called Zero Velocity Curves (ZVCs), i.e. the curves in the $(x,y)$ plane where the kinetic energy of the test particle vanishes. In Fig. \ref{conts}(a-d) we present the Hill's regions configurations when $r_S = 0.1$.

As it is seen in Fig. \ref{conts}, the specific value of the Jacobi constant dictates the shape of the corresponding Hill's regions configurations. More precisely, there are four distinct energy cases:
\begin{itemize}
  \item Energy Case I $(C > C_1)$: The Hill's region is divided into three disconnected regions, so the energetically permissible motion of the third body is either very close to the primaries or very far from them (see panel (a) of Fig. \ref{conts}).
  \item Energy Case II $(C_2 = C_3 < C < C_1)$: The neck around $L_1$ is open and the test particle can travel between the two primaries (see panel (b) of Fig. \ref{conts}).
  \item Energy Case III $(C_2 = C_3 < C < C_4)$: The necks around $L_2$ and $L_3$ opens and therefore a test particle orbiting around the primaries could escape from the system (see panel (c) of Fig. \ref{conts}).
  \item Energy Case V $(C < C_4)$: The energetically forbidden banana-shaped regions completely disappear.
\end{itemize}

It should be noted that in the pseudo-Newtonian case, $r_S > 0$, additional energetically forbidden circular regions emerge around the two primaries, where the radii of those circular regions coincide with the respective Schwarzschild radii. These additional circular forbidden regions, around each primary, are always present, regardless the specific value of the energy (or equivalently of the jacobi constant).

\section{Computational methodology}
\label{cometh}

For revealing the global orbital properties of the binary black hole system, we need to classify several sets of initial conditions of orbits. Our main numerical analysis will take place on the configuration $(x,y)$ plane. More precisely, for a given value of the Jacobi constant $C$, we choose a grid of $1024 \times 1024$ initial conditions for the position of the test particle, uniformly distributed in the area $-2.5 < x,y < + 2.5$. On the other hand, the initial velocities are set according to the conditions $\dot{r} = 0$ and $\dot{\phi} < 0$, where $(r,\phi)$ are the usual polar coordinates. Therefore, the initial conditions for the velocity in Cartesian coordinates are given by
\begin{align}
\dot{x_0} &= \frac{y_0}{r_0}g(x_0,y_0), \nonumber\\
\dot{y_0} &= - \frac{x_0}{r_0}g(x_0,y_0),
\label{vel}
\end{align}
where $r_0 = \sqrt{x_0^2 + y_0^2}$, while $g(x_0,y_0) = \sqrt{2 U(x_0,y_0) - C}$.

Each initial condition can be classified according to the type of motion described by the corresponding orbit. In particular we have:
\begin{itemize}
  \item Bounded orbits, which remain inside the scattering region after the end of the numerical integration.
  \item Orbits that escape from the system.
  \item Orbits which either collide with one of the primaries (when $r_S = 0$), or does get very close to the event horizon of them (when $r_S > 0$).
\end{itemize}

To perform such a classification, we consider a limiting circle with center at the origin $(0,0)$ and radius $R_d$. In the case where the test particle remains inside this circle after a sufficiently long time of the numerical integration $t_{\rm max}$, then we consider this motion as bounded. If the test particle on the other hand intersects, before $t = t_{\rm max}$, the limiting circle, with velocity pointing outwards, we have the case of escaping motion. It is evident that the definition of bounded and unbounded motion become more accurate as the values of $t_{\rm max}$ and $R_d$ are sufficiently large. In our calculations we choose $R_d = 10$, thus adopting the same value, used in the past in many similar cases \citep[e.g.,][]{N04,N05,Z15a,Z15b,Z15c}.

Additionally, it is worth noting that all bounded orbits can be further classified into regular an chaotic orbits. One of the simplest, yet very fast and accurate methods, is the Smaller Alignment Index (SALI) \citep{S01}, which is defined as
\begin{equation}
\rm SALI(t) \equiv min(d_{-}, d_{+}),
\label{sali}
\end{equation}
where
\begin{align}
d_{-} & \equiv \left\Vert\frac{{\vec{w_1}}(t)}{\| {\vec{w_1}}(t) \|} - \frac{{\vec{w_2}}(t)}{\| {\vec{w_2}}(t) \|}\right\Vert, \nonumber\\
d_{+} & \equiv \left\Vert\frac{{\vec{w_1}}(t)}{\| {\vec{w_1}}(t) \|} + \frac{{\vec{w_2}}(t)}{\| {\vec{w_2}}(t) \|}\right\Vert,
\label{align}
\end{align}
are the alignments indices, while ${\vec{w_1}}(t)$ and ${\vec{w_2}}(t)$, are two deviation vectors which initially are orthonormal and point in two random directions. At every time step, each deviation vector is normalized to 1. Computationally, each orbit is classified according to the final value of SALI at the end of the numerical integration. More precisely, according to \citet{SABV04}, if SALI $> 10^{-4}$ we have the case of a regular orbit, while if SALI $< 10^{-8}$ the orbit is chaotic. On the other hand, when $10^{-8} \leq \rm SALI \leq 10^{-4}$ we have the case of a sticky orbit\footnote{Sticky orbit refers to a special type of orbit which behaves as a regular one for a long-time interval and then it finally exhibits its true chaotic nature.}.

\begin{figure*}
\centering
\resizebox{\hsize}{!}{\includegraphics{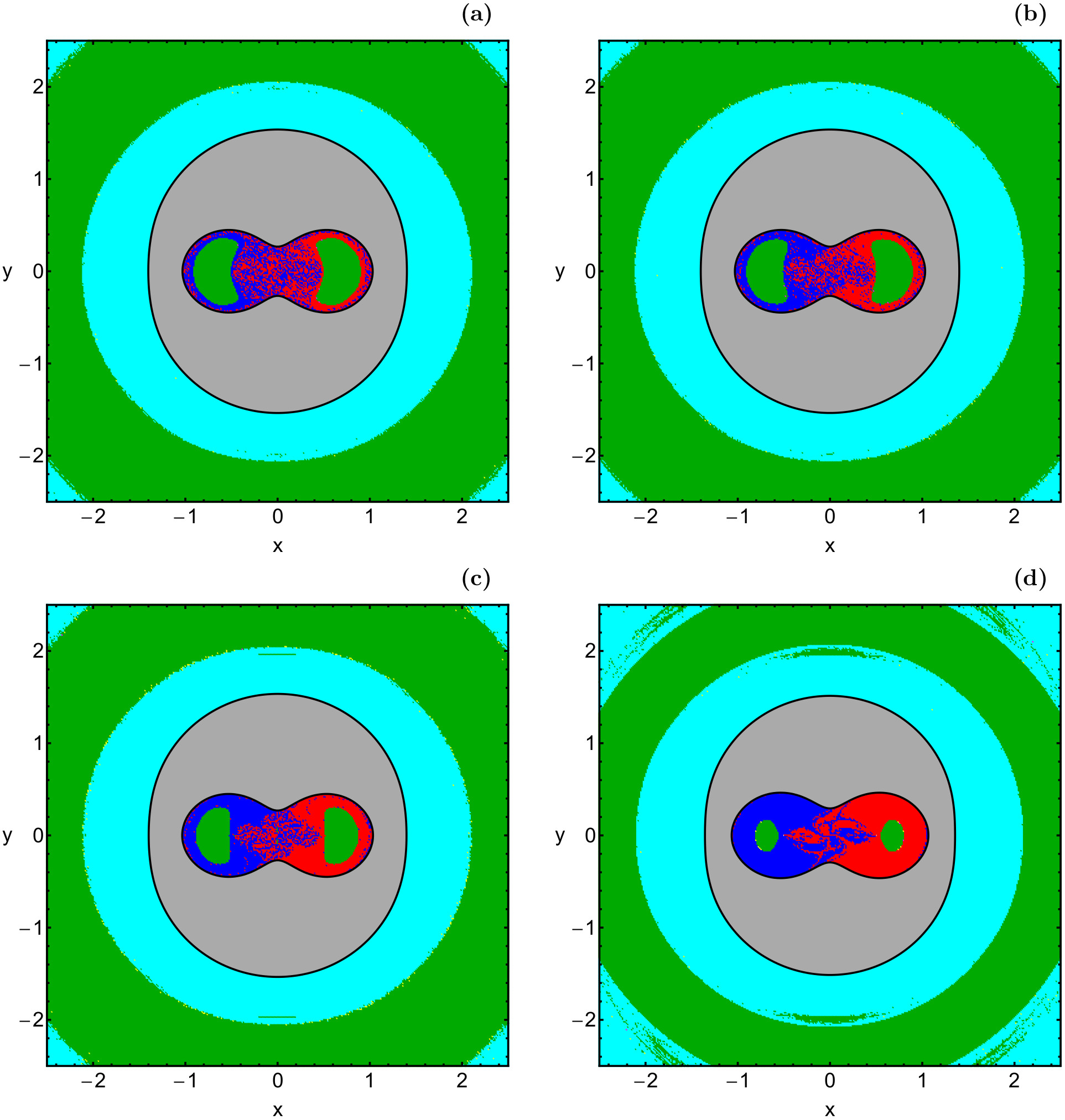}}
\caption{Basin diagrams on the configuration plane, for $C = 3.6$, when (a left): $r_S = 0$, (b): $r_S = 10^{-4}$, (c): $r_S = 10^{-3}$, and (d): $r_S = 10^{-2}$. The color code is the following: non-escaping regular orbits (green), trapped sticky orbits (magenta), trapped chaotic orbits (yellow), collision and close encounter orbits to primary 1 (blue), collision and close encounter orbits to primary 2 (red), and escaping orbits (cyan). The forbidden regions are shown in gray. (Color figure online).}
\label{c1}
\end{figure*}

\begin{figure*}
\centering
\resizebox{\hsize}{!}{\includegraphics{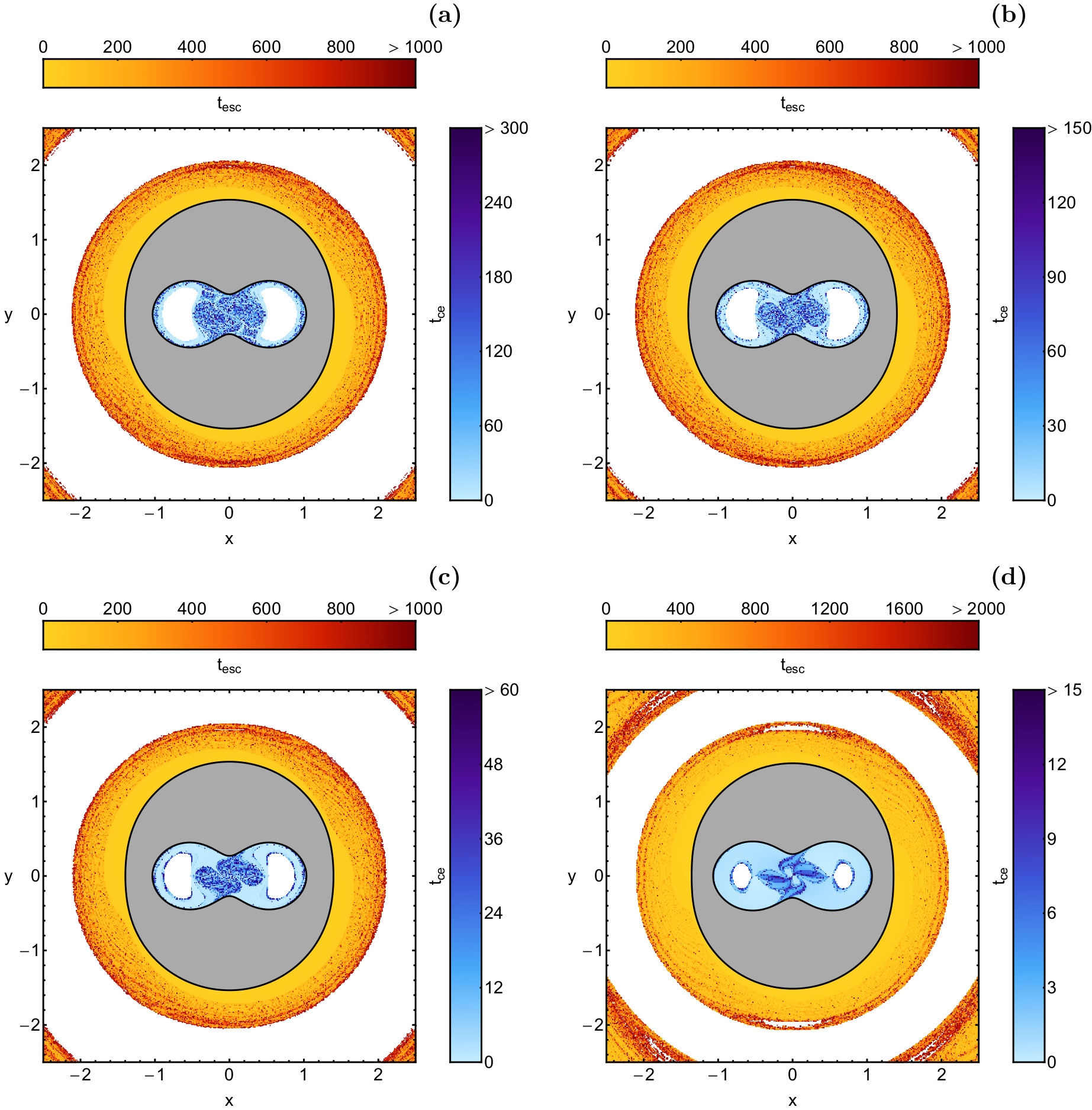}}
\caption{Distribution of times spent by the test particle in escape and collision/close encounter orbits. The white zones correspond to the initial conditions of bounded orbits (regular, sticky and chaotic). The initial conditions and all the parameters are as in Fig. \ref{c1}. (Color figure online).}
\label{c1t}
\end{figure*}

\begin{figure*}
\centering
\resizebox{\hsize}{!}{\includegraphics{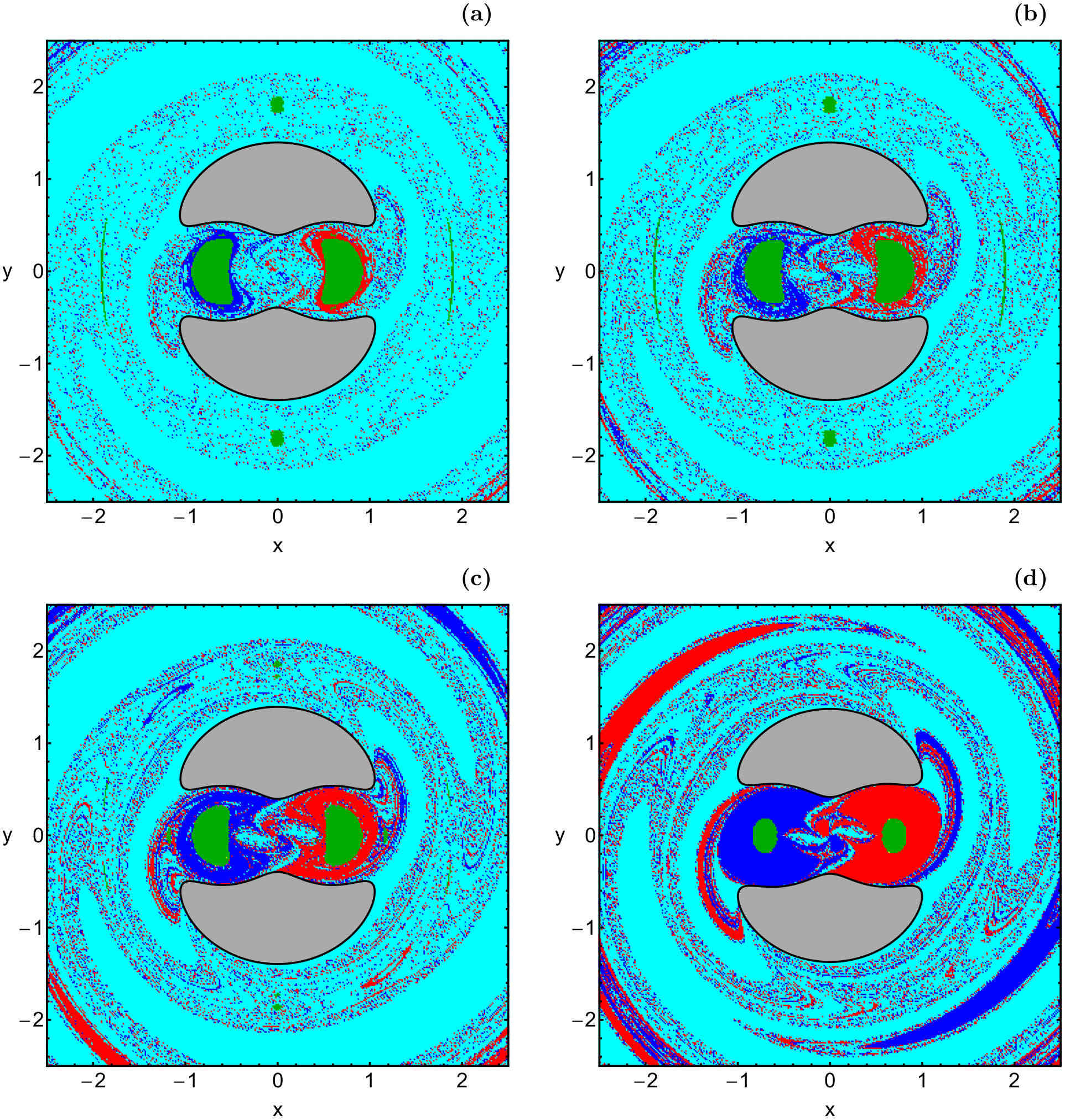}}
\caption{Basin diagrams of the configuration plane, for $C = 3.3$, when (a): $r_S = 0$, (b): $r_S = 10^{-4}$, (c): $r_S = 10^{-3}$, and (d): $r_S = 10^{-2}$. The color code is the same as in Fig. \ref{c1}. (Color figure online).}
\label{c2}
\end{figure*}

\begin{figure*}
\centering
\resizebox{\hsize}{!}{\includegraphics{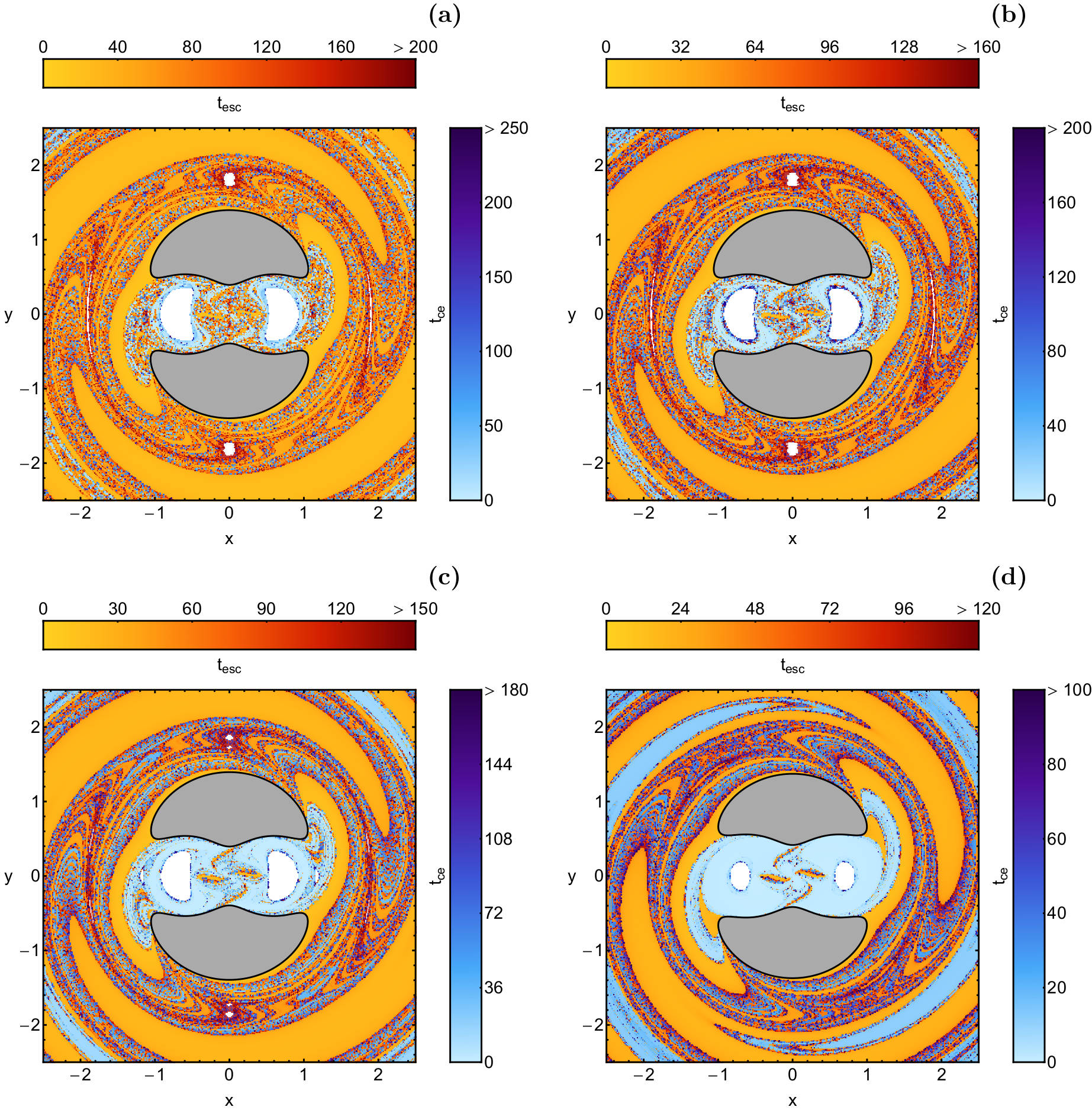}}
\caption{Distribution of times spent by the test particle in escape and collision/close encounter orbits. The white zones correspond to the initial conditions of bounded orbits (regular, sticky and chaotic). The initial conditions and all the parameters are as in Fig. \ref{c2}. (Color figure online).}
\label{c2t}
\end{figure*}

\begin{figure*}
\centering
\resizebox{\hsize}{!}{\includegraphics{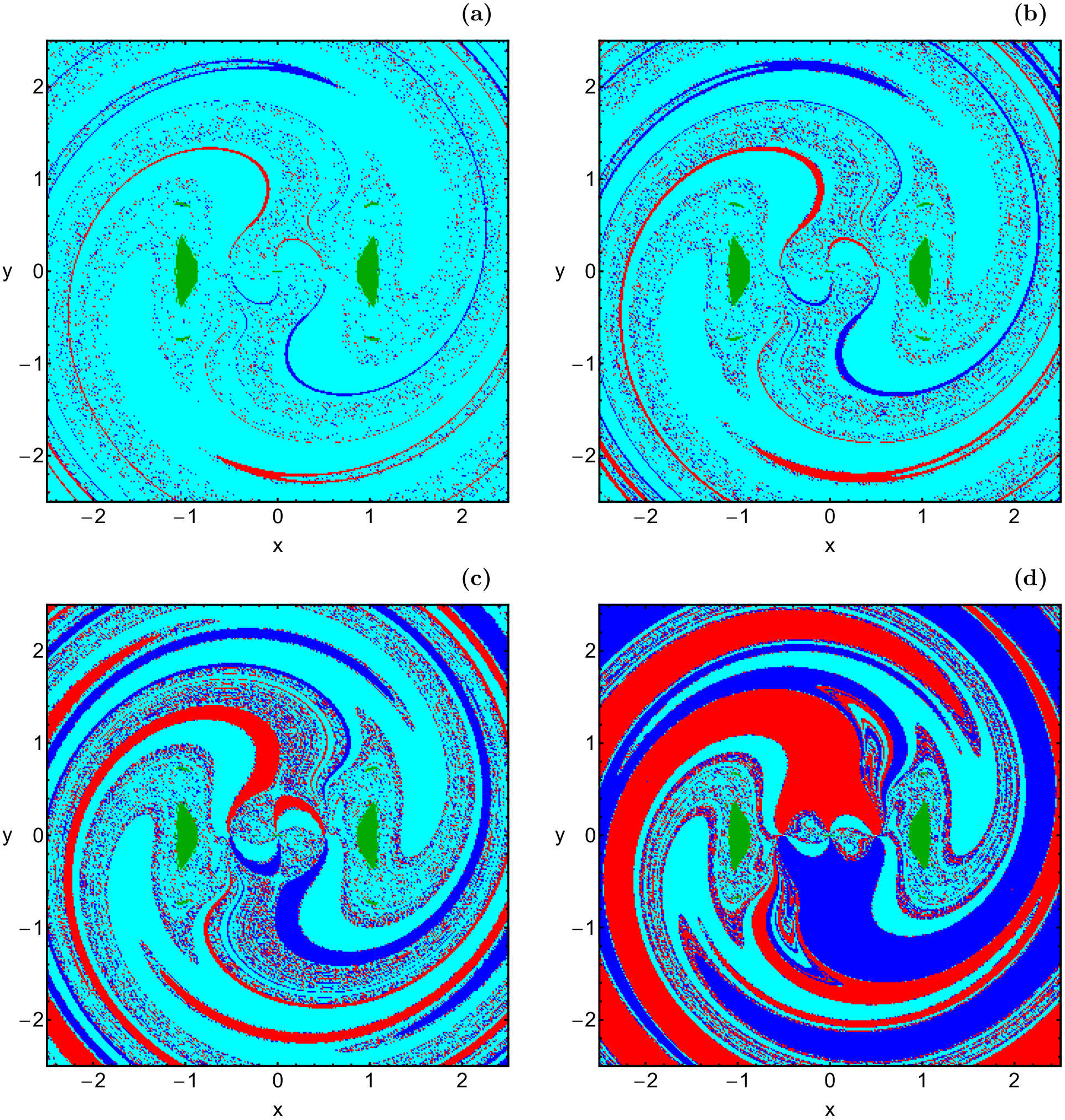}}
\caption{Basins diagrams on the configuration plane, for $C = 1.45$, when (a): $r_S = 0$, (b): $r_S = 10^{-4}$, (c): $r_S = 10^{-3}$, and (d): $r_S = 10^{-2}$. The color code is the same as in Fig. \ref{c1}. (Color figure online).}
\label{c4}
\end{figure*}

\begin{figure*}
\centering
\resizebox{\hsize}{!}{\includegraphics{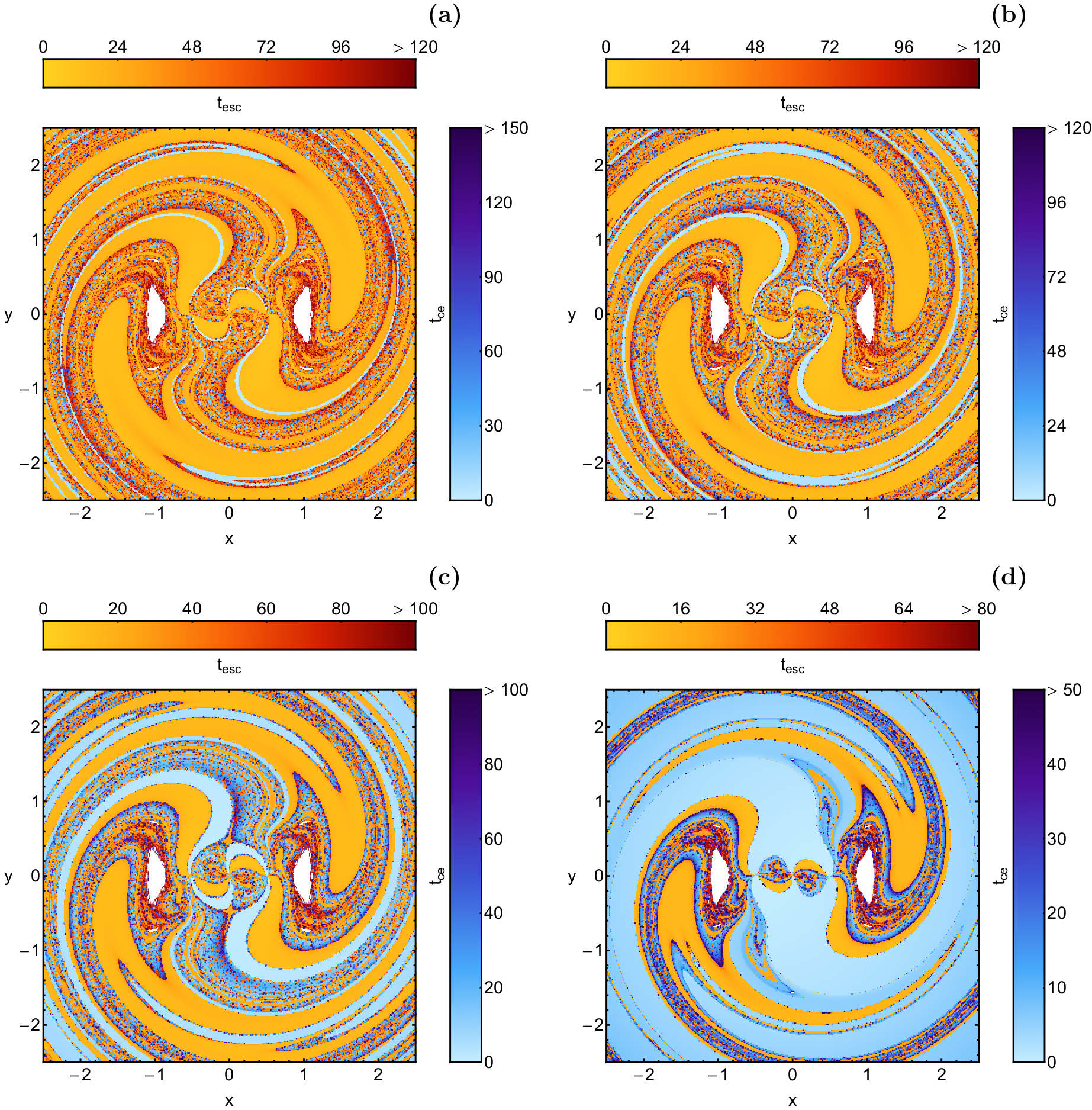}}
\caption{Distribution of times spent by the test particle in escape and collision/close encounter orbits. The white zones correspond to the initial conditions of bounded orbits (regular, sticky and chaotic). The initial conditions and all the parameters are as in Fig. \ref{c4}. (Color figure online).}
\label{c4t}
\end{figure*}

\begin{figure*}
\centering
\resizebox{\hsize}{!}{\includegraphics{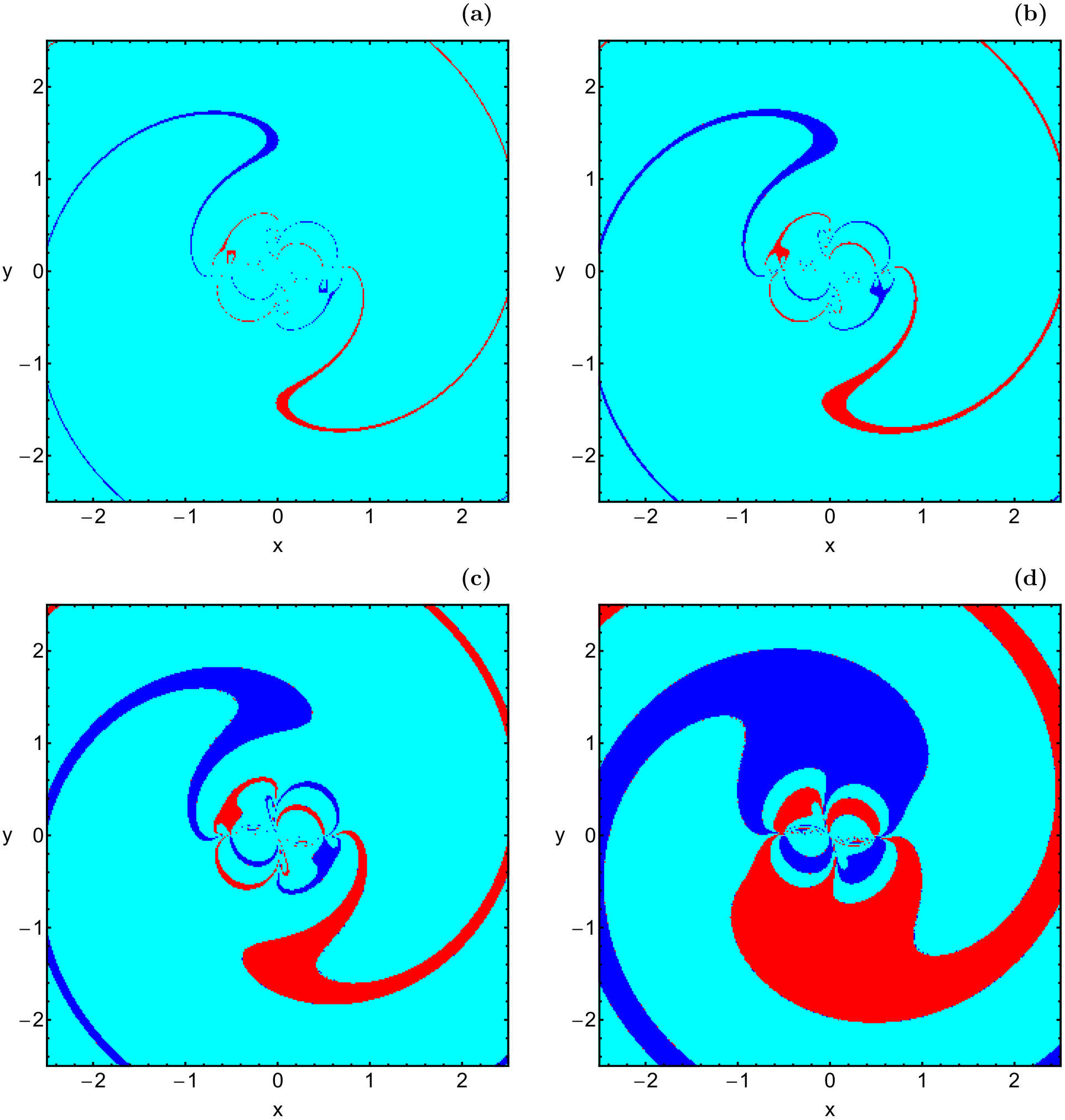}}
\caption{Basin diagrams on the configuration plane, for $C = -0.5$, when (a): $r_S = 0$, (b): $r_S = 10^{-4}$, (c): $r_S = 10^{-3}$, and (d): $r_S = 10^{-2}$. The color code is the same as in Fig. \ref{c1}. (Color figure online).}
\label{c5}
\end{figure*}

\begin{figure*}
\centering
\resizebox{\hsize}{!}{\includegraphics{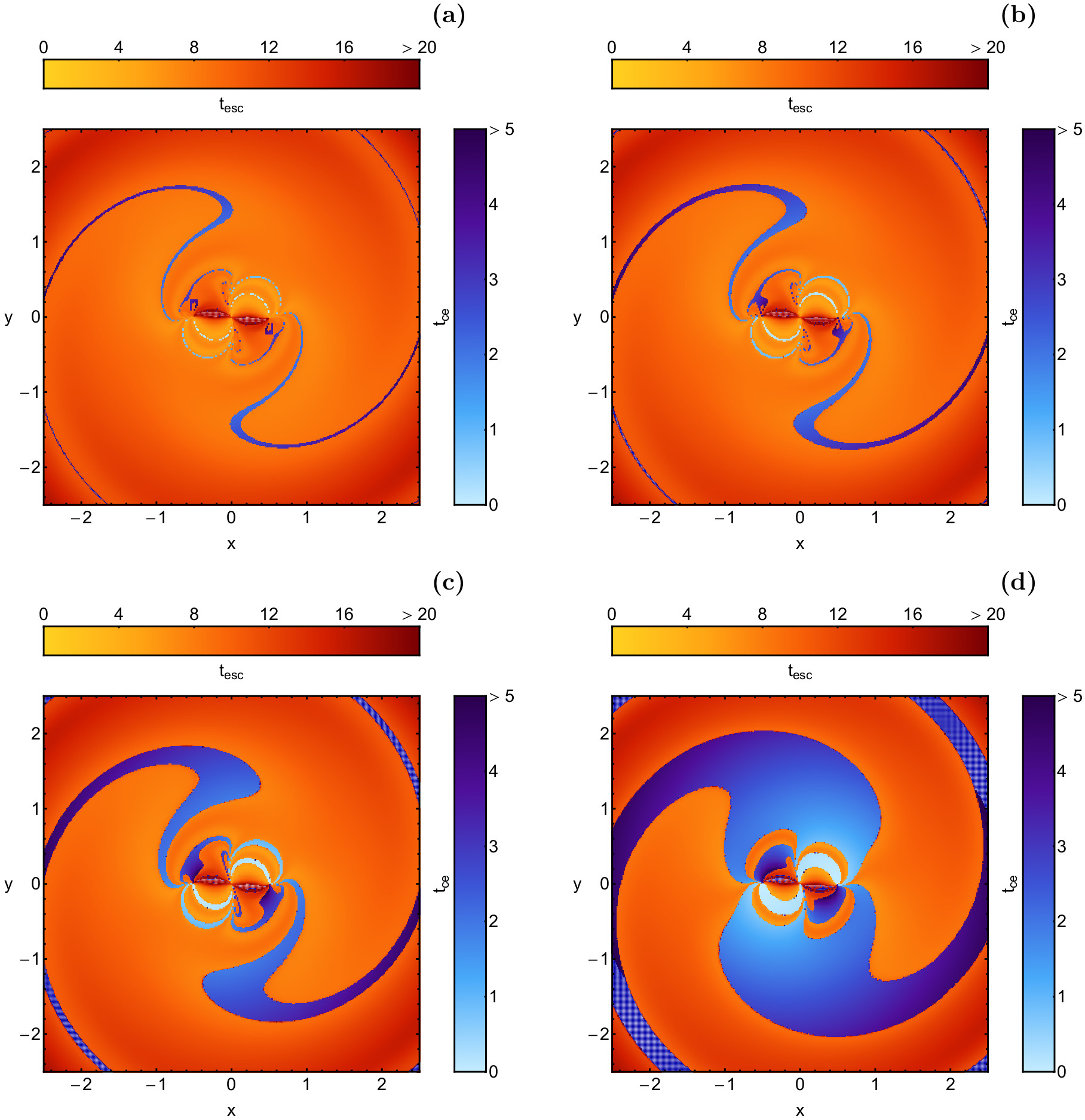}}
\caption{Distribution of times spent by the test particle in escape and collision/close encounter orbits. The white zones correspond to the initial conditions of bounded orbits (regular, sticky and chaotic). The initial conditions and all the parameters are as in Fig. \ref{c5}. (Color figure online).}
\label{c5t}
\end{figure*}

\begin{figure*}
\centering
\resizebox{\hsize}{!}{\includegraphics{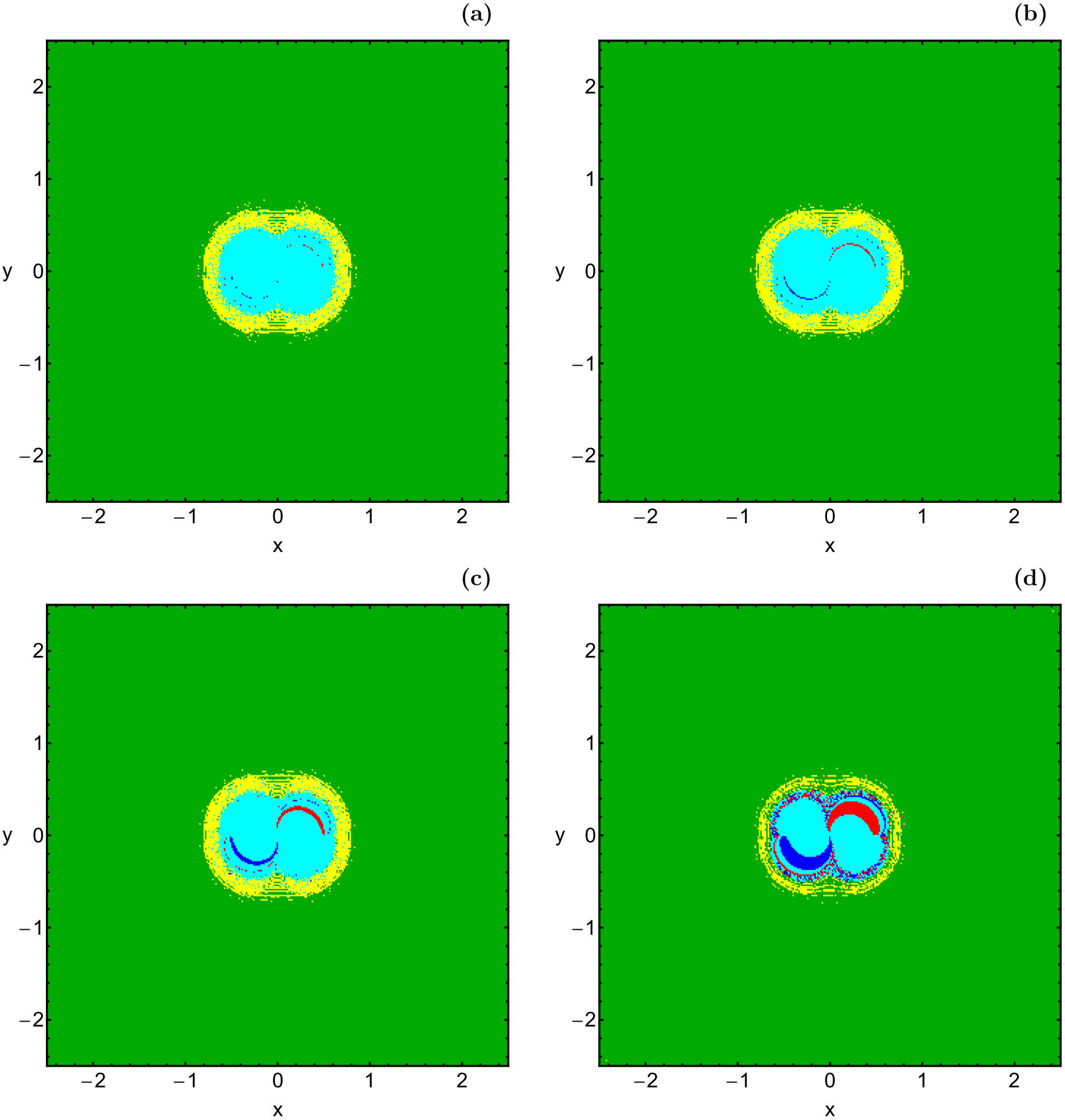}}
\caption{Basin diagrams on the configuration plane, for $C = -2$, when (a): $r_S = 0$, (b): $r_S = 10^{-4}$, (c): $r_S = 10^{-3}$, and (d): $r_S = 10^{-2}$. The color code is the same as in Fig. \ref{c1}. (Color figure online).}
\label{c6}
\end{figure*}

\begin{figure*}
\centering
\resizebox{\hsize}{!}{\includegraphics{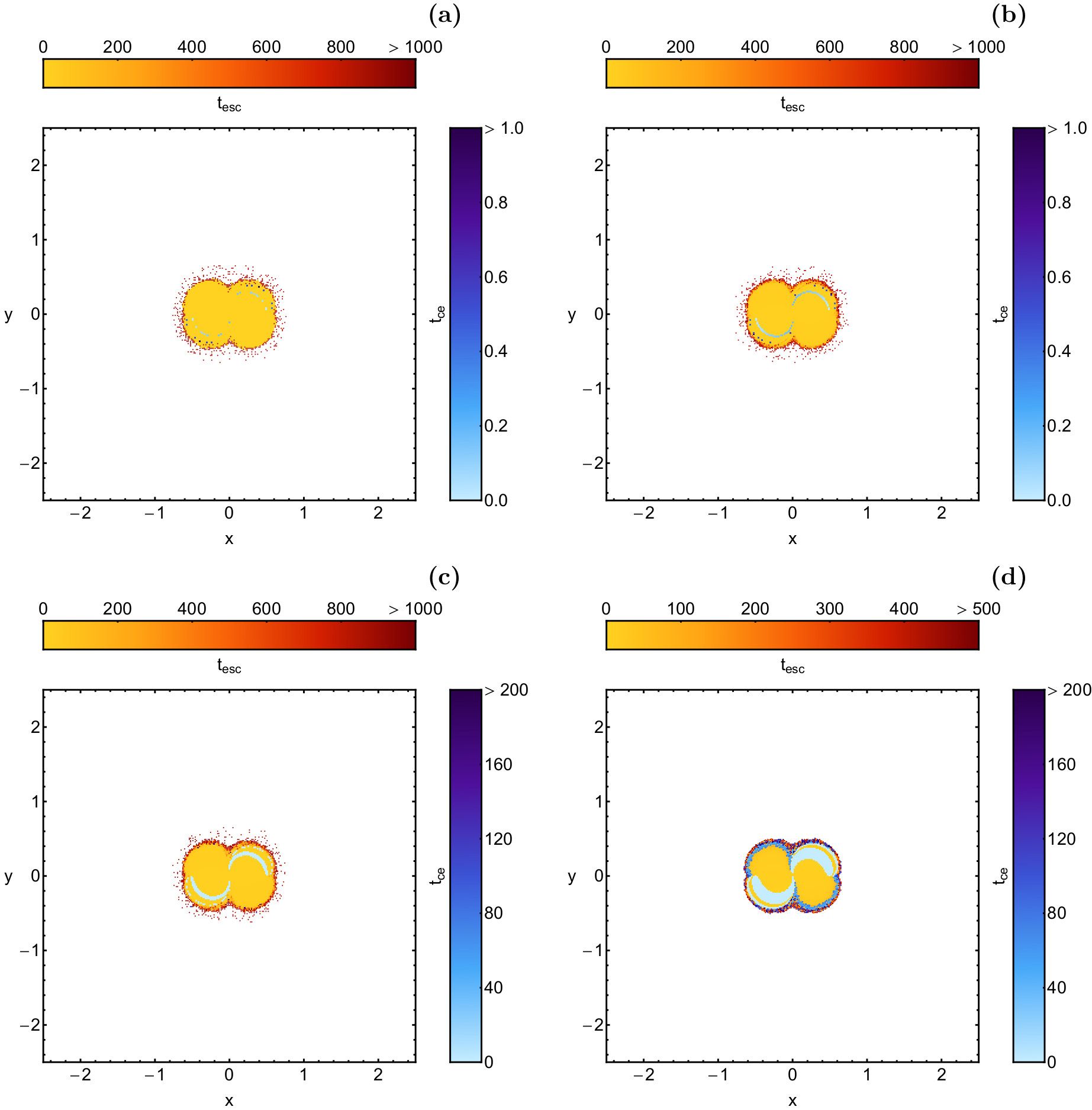}}
\caption{Distribution of times spent by the test particle in escape and collision/close encounter orbits. The white zones correspond to the initial conditions of bounded orbits (regular, sticky and chaotic). The initial conditions and all the parameters are as in Fig. \ref{c6}. (Color figure online).}
\label{c6t}
\end{figure*}

\begin{figure*}
\centering
\resizebox{\hsize}{!}{\includegraphics{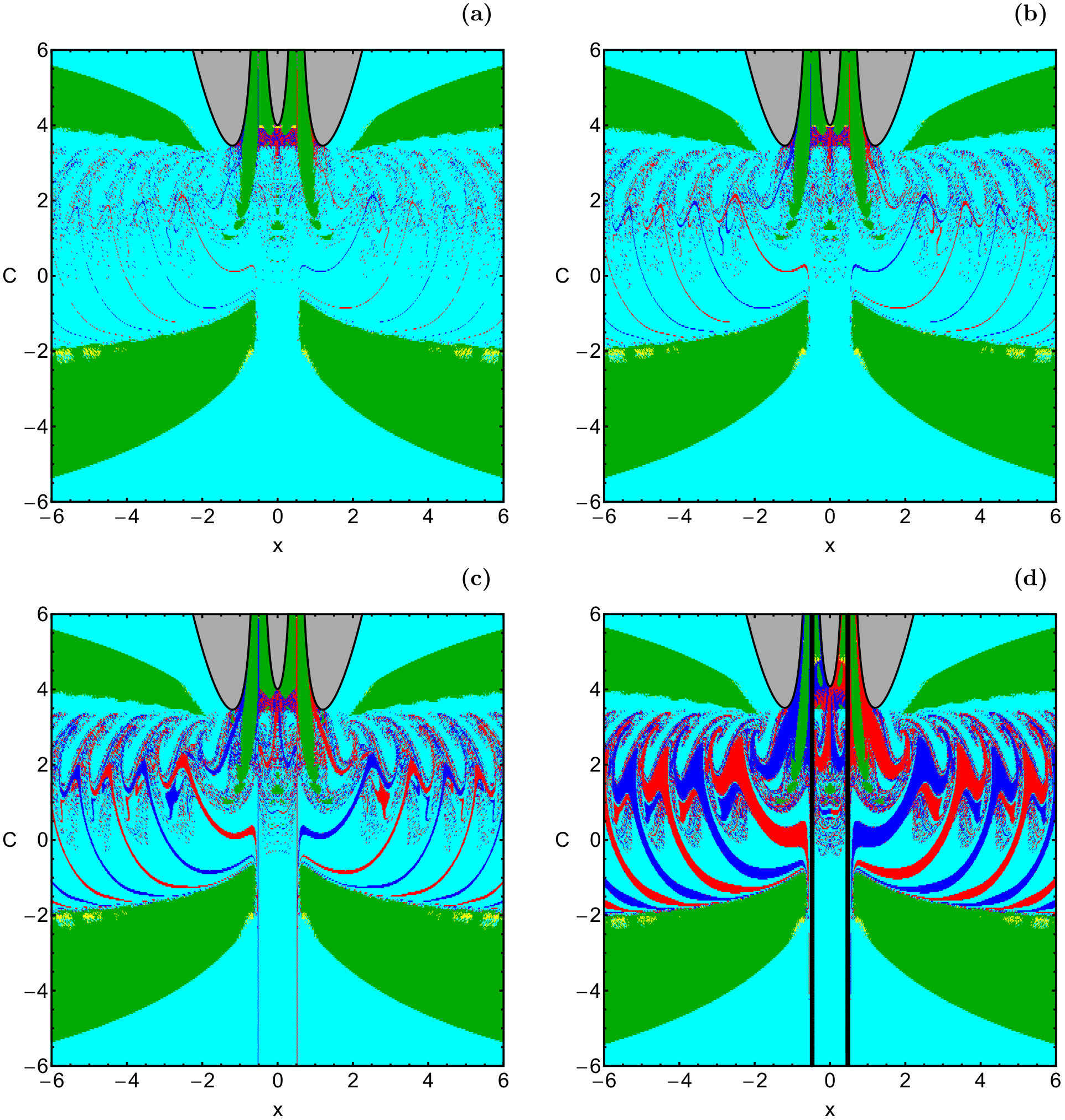}}
\caption{Basin diagrams on the $(x,C)$ plane, when (a): $r_S = 0$, (b): $r_S = 10^{-4}$, (c): $r_S = 10^{-3}$, and (d): $r_S = 10^{-2}$. The color code is the same as in Fig. \ref{c1}. (Color figure online).}
\label{xc}
\end{figure*}

\begin{figure*}
\centering
\resizebox{\hsize}{!}{\includegraphics{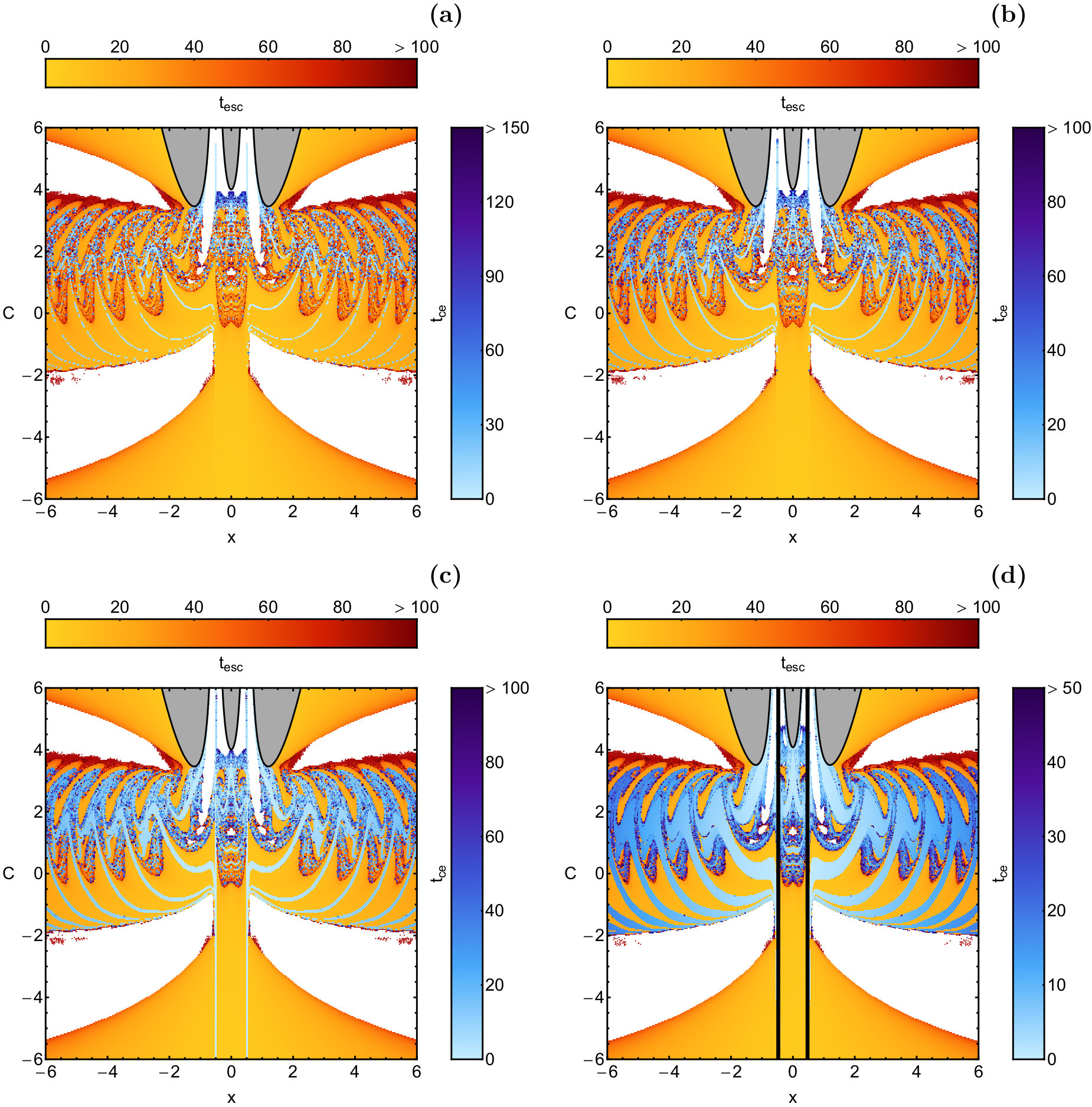}}
\caption{Distribution of times spent by the test particle in escape and collision/close encounter orbits. The white zones correspond to the initial conditions of bounded orbits (regular, sticky and chaotic). The initial conditions and all the parameters are as in Fig. \ref{xc}. (Color figure online).}
\label{xct}
\end{figure*}

\begin{figure*}
\centering
\resizebox{\hsize}{!}{\includegraphics{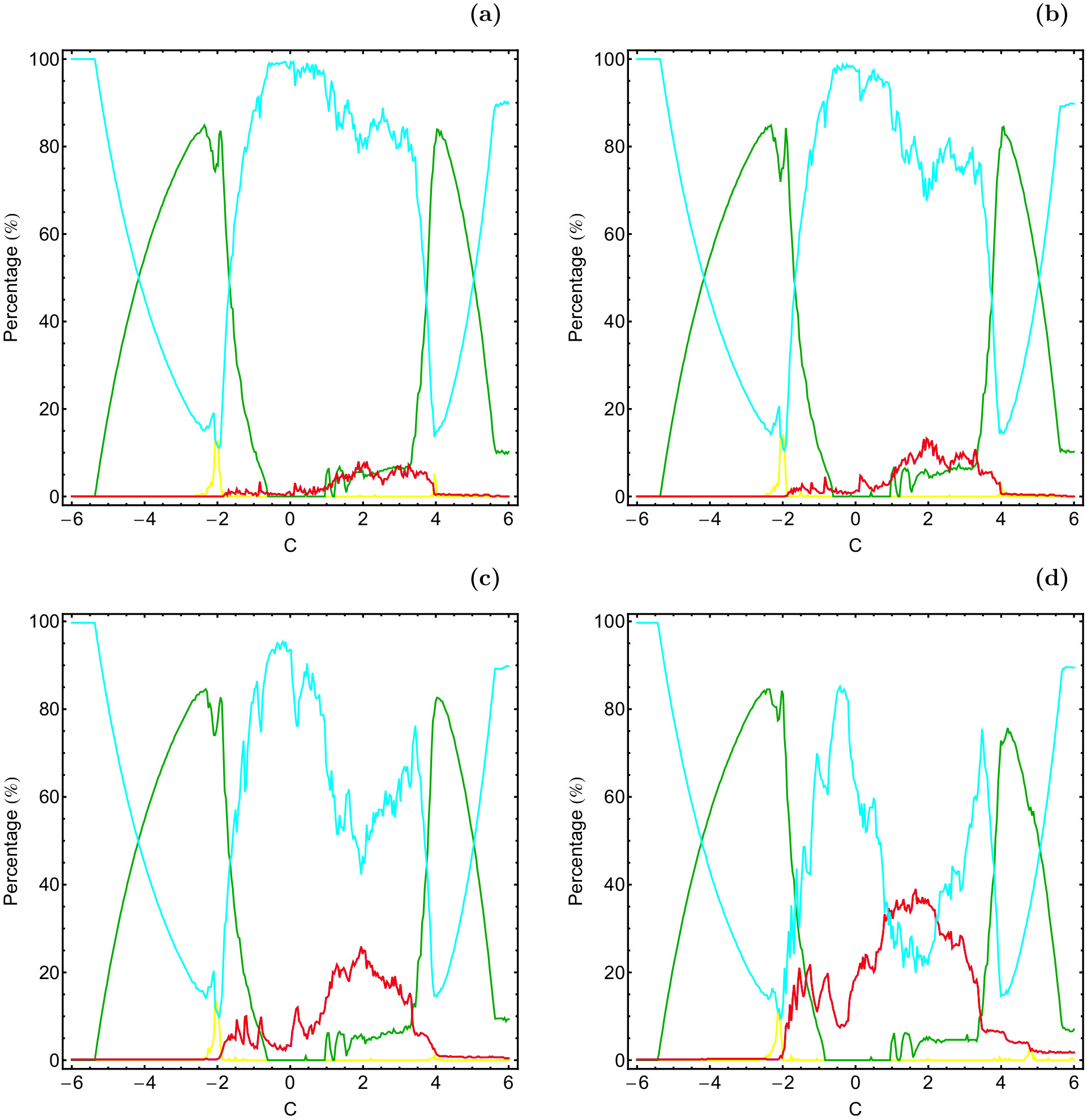}}
\caption{Percentages of escaping (cyan), regular (green) and collision/close encounter orbits (red), in terms of the Jacobi constant, when (a): $r_S = 0$, (b): $r_S = 10^{-4}$, (c): $r_S = 10^{-3}$, and (d): $r_S = 10^{-2}$. Color figure online).}
\label{pxc}
\end{figure*}

\begin{figure*}
\centering
\resizebox{\hsize}{!}{\includegraphics{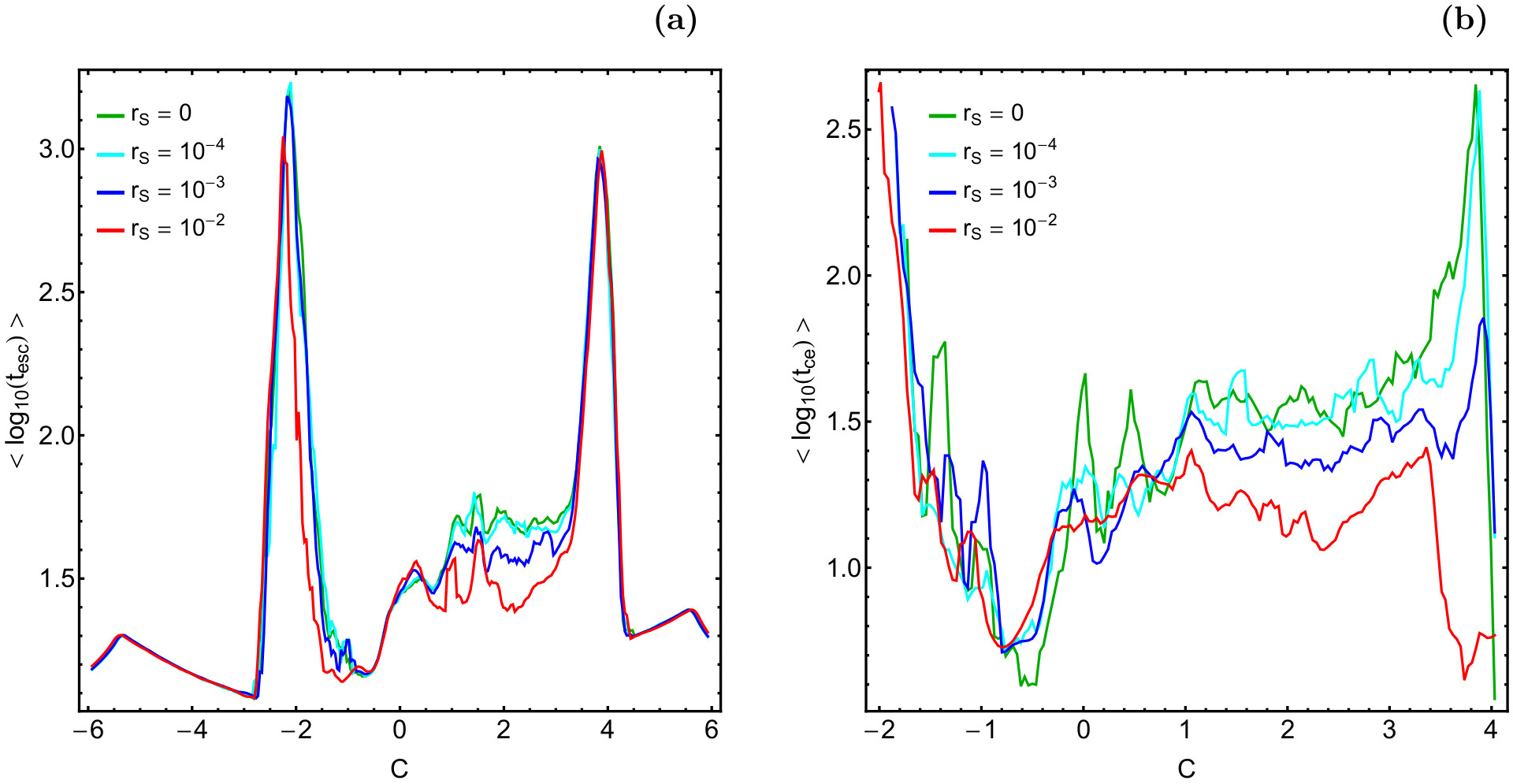}}
\caption{Evolution of the average logarithmic (a): escape and (b): collision/close encounter time $(< \log_{10}(t) >)$ of the orbits on the $(x,C)$ plane, in terms of the Jacobi constant $C$. (Color figure online).}
\label{times}
\end{figure*}

\begin{figure}
\centering
\resizebox{\hsize}{!}{\includegraphics{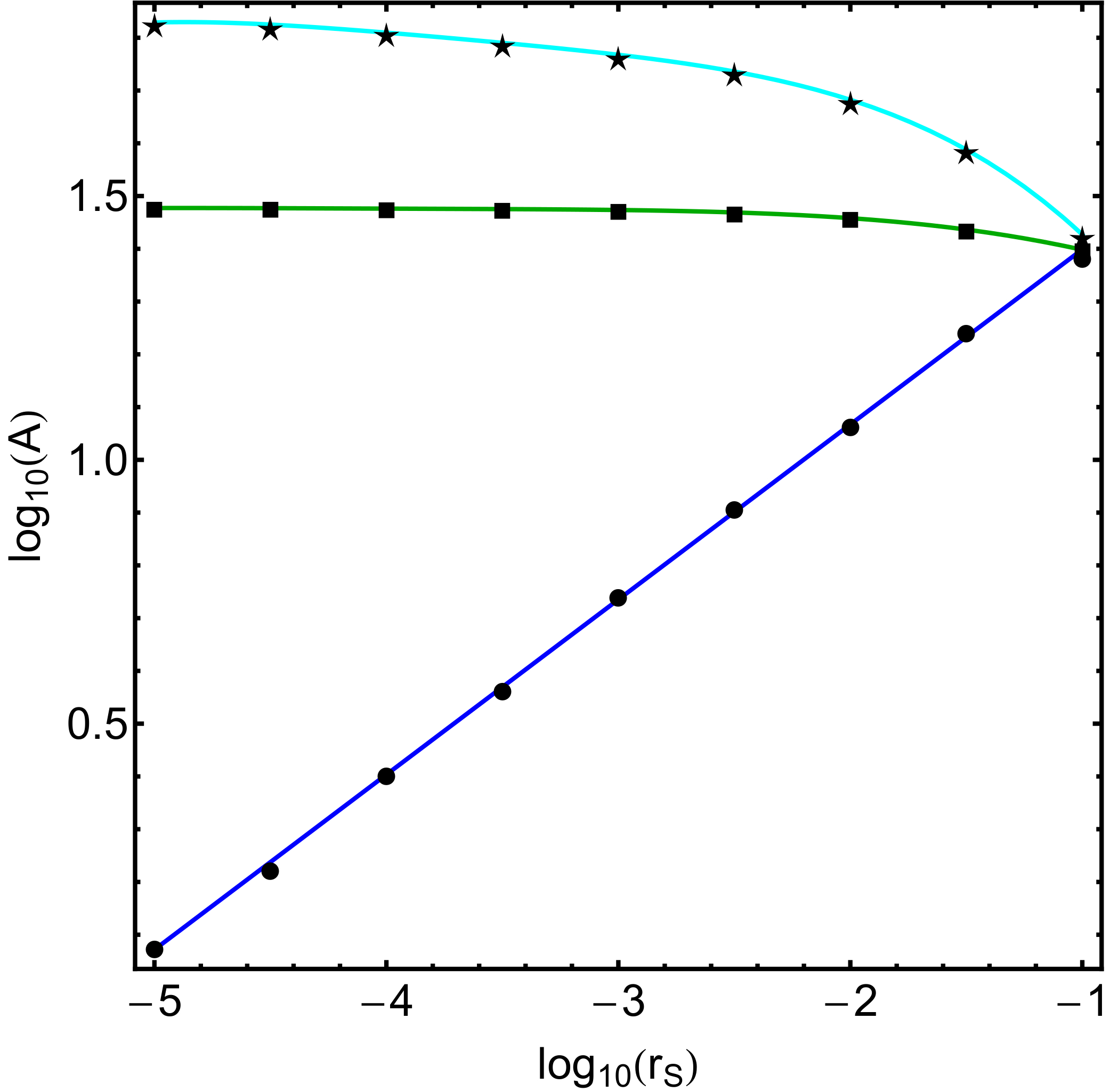}}
\caption{Evolution of the area of the different types (bounded, escape and close encounter) of basins on the $(x,C)$ plane, in terms of the Jacobi constant. The color code for distinguishing the types of the orbits is the same used in Fig. \ref{c1}. (Color figure online).}
\label{rs}
\end{figure}

As we have seen in the previous section, in the pseudo-Newtonian case, that is when $r_S > 0$, there is an additional energetically forbidden region around each primary, where each surface plays the role of the event horizon. Therefore, a test particle cannot enter the event horizon, but it can move sufficiently close to it. In this case we assume that we have the phenomenon of a ``close encounter". It is well known that all types of numerical integrators are extremely ill-behaved (e.g., zero time step) near the vicinity of a singularity. In order to avoid all such malfunctions we define a close encounter radius $r_{\rm ce} = 10^{-5}$. Moreover, the numerical integration is stopped when the test particle crosses the disk with radius $R_c = r_S + r_{\rm ce}$, around each black hole, with velocity pointing inwards. In other words, the numerical integration is stopped, just before the emergence of all the unwanted numerical issues. The fate of all the test particles that enter the disk with radius $R_c$ must be to be eaten by the black holes, since neither light nor particles can escape through this region from inside. In the classical Newtonian case, where $r_S = 0$, the test particle can, theoretically, approach infinitely close to the primaries. However, it is computationally impossible to follow the space evolution of an orbit infinitely close to the primaries. Our previous experience suggests that the integrator can properly work up to a radius of $10^{-5}$ around the primaries \citep[see e.g.,][]{N04,N05}. Therefore, if a test particle crosses the disk with radius $10^{-5}$, around one primary, with velocity pointing inwards, we assume that a collision takes place.

In all cases, the maximum integration time is set to $10^4$ time units. We choose to use such a high value of $t_{\rm max}$ in order to be sure that all the orbits have enough time for revealing their true nature.

For the numerical integration of the equations of motion (\ref{eqmot}) and the corresponding variational equations (\ref{variac}), we used a double precision Bulirsch-Stoer \verb!FORTRAN 77! algorithm \citep{PTVF92}. In all our calculations the numerical error, related to the conservation of the Jacobi integral (\ref{ham}) was, in most of the cases, smaller than $10^{-14}$. All the graphical illustration has been created using the latest version 11.2 of the software Mathematica$^{\circledR}$ \citep{Wolf03}.

\section{Orbit classification}
\label{orbclas}

For the classification of the initial conditions of orbits on the $\dot{\phi} < 0$ part of the surface of section $\dot{r} = 0$, we will follow the approach introduced in \citet{N04} and \citet{N05}. In particular, we will use color-coded diagrams (CCDs) in which each pixel is assigned a specific color, according to the particular type of the orbit. We may argue, that these CCDs are in fact a modern version of the classical Poincar\'{e} Surface of Section. These diagrams will help us locate the several types of basins\footnote{In this article, the term ``basin" is understood as the set of initial conditions leading to the same final state (bounded motion, escape or close encounter.)} on the configuration $(x,y)$ plane.

In the following subsections we will consider three cases, regarding the energy level (which is equivalent to the Jacobi constant). In each case, the geometry of the Hill's regions configurations changes drastically. At this point, we should clarify that the case $C > C_1$ is not taken into consideration mainly for saving space for other cases with much more interesting orbital structure.

\subsection{Energy case I $(C_1 < C < C_2 = C_3)$}
\label{ss1}

Let us start considering the case where the neck around the Lagrange point $L_1$ is open, such that the third body can freely travel around the primaries in the interior region. The basin diagrams when $C = 3.6$, are presented in Fig. \ref{c1}(a-d), for four values of the Schwarzschild radius $r_S$. In all cases, we see that inside the interior region there are two stability islands, which correspond to regular orbits around each primary. The bounded basins are surrounded by an entangled mixture of initial conditions, corresponding to orbits that either collide with the primaries (when $r_S = 0$) or exhibit the phenomenon of close encounter (when $r_S > 0$). As expected, the exterior region is dominated by bounded and escaping orbits. In particular, it is observed that the regular initial conditions form an annulus-shaped basin. Further numerical calculations suggest that all the initial conditions of the annulus correspond to simple 1:1 loop orbits moving around both primaries.

With increasing value of the Schwarzschild radius the following interesting phenomena take place:
\begin{itemize}
  \item The covered area by the regular islands in the interior region gradually decreases.
  \item The thickness of the annulus of bounded orbits located in the exterior region gets reduced.
  \item When $r_S = 10^{-2}$ (see panel (d) of Fig. \ref{c1}) additional small islands of regular motion emerge in the exterior region. These initial conditions correspond to secondary resonant 3:3 orbits moving around both primaries.
  \item The mixture of the close encounter orbits in the interior region become less chaotic. In other words, we may argue that the fractality\footnote{We would like to note that when we use the term ``fractal" for describing an area it simply means that this particular local region displays a fractal-like geometry, however without conducting any specific calculations as in \citet{AVS01,AVS09}.} of the interior regions is reduced.
\end{itemize}

The distribution of times spent by a test particle in escape (using tones of red) and collision/close encounter (using tones of blue) orbits are illustrated in panels of Fig. \ref{c1t}. It is interesting to note that the highest values of the collision and close encounter times correspond to initial conditions situated inside the fractal areas of the interior region. Similarly, the highest values of the escape times are associated with initial conditions in the neighborhood of the boundaries of the basins of escape.

\subsection{Energy case II $(C_2 = C_3 < C < C_4)$}
\label{ss2}

We continue with the case in which the region of non-allowed motion around $L_2$ and $L_3$ disappears. This means that the test particle can freely move between the interior and the exterior regions. In Fig. \ref{c2}(a-d) we present basin diagrams on the $(x,y)$ plane, when $C = 3.3$. One may observe that bounded regular motion is still possible, mainly around each primary. Indeed, two stability islands are still present very close to the two primary bodies. On the other hand, the presence of regular orbits, that circulate around both primaries, is either extremely weak or even impossible. The exterior region is covered both by fractal regions and well-formed basins of escape and collision/close encounter, which form complicated spiral structures.

As the value of the Schwarzschild radius increases, there are three important aspects that take place on the $(x,y)$ plane:
\begin{itemize}
  \item As in the energy case I, the area of the regular islands situated in the interior region gradually decreases.
  \item The covered area by the regular islands, situated in the exterior region also decreases, while in the case $r_S = 10^{-2}$, there are no sings of regular bounded motion around both primaries.
  \item The area of the basins of close encounters, mainly around the stability islands, increases significantly.
\end{itemize}

In Fig. \ref{c2t} we show the distribution of times spent by the test particle in escape and collision/close encounter orbits. Looking at the range of values at the two accompanying color bars, we observe that both the escape and the close encounter times of the orbits decrease with increasing value of the Schwarzschild radius.

\subsection{Energy case III $(C < C_4)$}
\label{ss3}

The last case concerns the scenario where the whole $(x,y)$ plane is available for the test particle since the forbidden regions disappear. Our analysis suggests that this energy range $(C < C_4)$ is particularly interesting from a dynamical point of view due to the fact that the orbital structure of the $(x,y)$ plane varies drastically with the changes in the Jacobi constant $C$.

There are three different types of geometry for the basin diagrams on the $(x,y)$ plane. The first type is presented in Fig. \ref{c4}, where $C = 1.45$. It is seen that in this case, when $r_S = 0$ (panel (a)), the configuration plane is covered by a complex blend of escape and collision orbits. However as we proceed to the pseudo-Newtonian cases, with $r_S > 0$, the fractal areas are heavily reduced, thus giving place to well-formed basins of escape and mainly to basins of close encounters. Non-escaping regular motion, around each primary, is still possible, while we may argue that the shift on the Schwarzschild radius does not practically affect the size of the bounded basins. The corresponding distributions of the escape, as well as the collision/close encounter times of the orbits, are illustrated in Fig. \ref{c4t}.

In Fig. \ref{c5}(a-d) we observe the second possible geometry of the $(x,y)$ plane when $C = -0.5$. It is evident that this case is completely different, in comparison with the results of Fig. \ref{c4}. The entire plane is dominated by well-formed basins (of escape and collision/close encounters), while the most remarkable feature is the almost complete absence of fractal regions. Once more, the area of the basins of the close encounter significantly increases, as the Schwarzschild radius grows. Furthermore, it should be noted that bounded regular motion is not possible in this energy region. In Fig. \ref{c5}(a-d) we show the corresponding distributions of the escape and the collision/close encounter times of the orbits. Interestingly enough, the increase of the Schwarzschild radius seems not to affect the distributions of times.

The third possible type of geometry is illustrated in Fig. \ref{c6}(a-b), where $C = -2$. Now the basins pattern completely changes, a high percentage of the $(x,y)$ plane is dominated by bounded regular orbits that circulate around both primaries. On the contrary, for such high energy values, there are not regular orbits moving around one of the primary bodies. Near the central region, we detect a layer composed of trapped chaotic orbits, while inside this layer we observe the presence of escaping and close encounter initial conditions. Two new phenomena take place with the increase of the Schwarzschild radius: (i) the layer of trapped chaotic orbits weakens and (ii) the area of the basins of the close encounter orbits increases, while at the same time the size of the basins of escape is reduced. The distribution of times spent by the third particle in escape and collision/close encounter orbits are shown in Fig. \ref{c6t}(a-d).

In Fig. \ref{c6} we have seen that the basins of non-escaping regular orbits cover most of the $(x,y)$ plane. Therefore, the natural question that arises is the following: do the basins of regular orbits extend to infinity for such high values of the energy (or equivalently, small values of the Jacobi constant)? To answer this question we expanded our scattering region to $-7 \leq x, y \leq +7$ and we repeated the orbit classification. We found that the basins of bounded (both regular and chaotic) orbits extend up to radius $R = \sqrt{x^2 + y^2} < 5.5$. More specifically, in the region $4 < R < 5.5$ there is a constant transition between ordered and chaotic orbits. For larger values of the radius $R$, we find that bounded orbits disappear and the whole plane is dominated, first by a fractal mixture and then by basins of escape as well as of basins of close encounters. Our analysis suggests that both basins of escape and close encounters extend to infinity.

\subsection{An overview analysis}
\label{over}

All the CCDs presented in the previous subsections (see Figs. \ref{c1}, \ref{c2}, \ref{c4}, \ref{c5}, and \ref{c6}), reveal the orbital structure of the configuration $(x,y)$ plane for a fixed value of the Jacobi constant (or equivalently, for a fixed orbital energy) in the case of retrograde motion $(\dot{\phi} < 0)$. In order to obtain the same information for a continuous interval of $C$ values, we define a plane of representation in which the $x$-coordinate is the abscissa and the ordinate is given by the Jacobi constant. This means that all orbits start with initial position $(x_0, 0) $, i.e, on the $x$-axis, with initial velocities $(0,\dot{y}_{0})$, where the $y$-component of the velocity is derived from Eq. (\ref{vel}).

In Fig. 14 we present the basin diagram on the $(x,C)$ plane for $C \in [ˆ-6, 6]$, using four different values of the Schwarzschild radius. The black continuous line indicates the corresponding Zero Velocity Curve (ZVC), and it is defined as
\begin{equation}
f(x,C) = 2U(x,y = 0) = C.
\label{zvc}
\end{equation}

In all cases, zones of bounded regular motion around each primary exist mainly between the energetically forbidden regions. Furthermore, far from the primaries, we observe four stability islands, corresponding to regular orbits around both primary bodies. Between these four stability islands, one may observe a very complicated orbital structure composed of a unified basin of escape along with several basins of collision and close encounters. With increasing value of the Schwarzschild radius $r_S$ the area of the basins of close encounters grows, while the area covered by the stability basins keeps unaffected. In Fig. \ref{xct} we present the distribution of times employed by the test particle in the escape and collision/close encounter orbits. Once again, it can be noted that the longest times are associated with initial conditions placed in the fractal basin.

In order to obtain a complete view, regarding the evolution of the orbital dynamics of the system, we computed the percentage of orbits leading to the different types of orbits (escaping, regular, chaotic and collision orbits or close encounter), in terms of the Jacobi constant and using four values of the Schwarzschild radius. From Fig. \ref{pxc} it is manifested that escaping motion dominates for high values (positive and negative) of $C$, i.e., $C < - 4.5$ and $C > 5$. More precisely, for $C < - 5.5$, the escaping orbits are the unique type of orbits that survive. In addition, the parametric evolution of the percentage of non-escaping regular orbits is almost the same in all four cases, leading to the conclusion that the value of the Schwarzschild radius has almost zero influence in this type of orbits. On the contrary, the rate of escaping and collision/close encounter orbits is highly dependent of the Schwarzschild radius. In particular, in the energy region $- 2.5 < C < 4$, it can be observed that with increasing the value of $r_S$ the percentage of escaping orbits is reduced, while the percentage of close encounter orbits rises. The percentage of trapped chaotic orbits has an unchanging behavior, with null values except for $C \approx -ˆ'2$ where the percentage reaches a peak of about 10\%.

Additional information could be extracted from the escape and collision/close encounter times. In Fig. \ref{times} we plot the evolution of the average logarithmic value of the escape and collision/close encounter times $(< \log_{10}(t) >)$, in terms of the Jacobi constant. In panel (a) it is seen that, for a large set of $C$ values, the average escape time of the orbits is almost the same for all values of the Schwarzschild radius $r_S$. However in the range $1 < C < 3$ we observe that the average escape time of the orbits is reduced, with increasing value of $r_S$. Interestingly the same also applies to the average close/encounter time, given in panel (b) of Fig. \ref{times}. Indeed in the range $1 < C < 4$, one can identify a clear pattern of reduction of $< \log_{10}(t_{\rm ce}) >$ in relation to $r_S$. Therefore we may conclude that the Schwarzschild radius $r_S$ does locally (in a specific energy range) influence the average escape and close/encounter time of the orbits.

Before closing this section, it is worth to discuss the relation between the area of the different types of basins and the Schwarzschild radius $r_S$. In Fig. \ref{rs} we depict the evolution of the area of the different types of basins on the $(x,C)$ plane, as a function of the Schwarzschild radius $r_S$. We observe that the area of the non-escaping regular orbits remains almost unperturbed and only for high values of $r_S$ shows a small reduction. Moreover, the area of the basins of escape is constantly reduced with the increase in the value of the Schwarzschild radius. Finally, the most interesting behavior is that of the close encounter orbits, where it is seen that the area of the corresponding basins exhibits a perfect linear increase. At this point, it should be noted that in \citet{N04} (see Fig. 6) and \citet{N05} (see Fig. 11) a very similar linear increase it was observed, regarding the area of the basins of collision and the corresponding radius of collision. When $r_S = 10^{-1}$ close encounter orbits (to both primaries) occupy half of the $(x,C)$ plane, while non-escaping regular and escaping orbits share the rest half of the same plane.

\section{Concluding remarks}
\label{conc}

The orbital dynamics of a test particle in the presence of an equal-mass non-spinning binary black hole system has been numerically investigated in the context of the pseudo-Newtonian circular restricted three-body problem. After numerically integrating several large sets of initial conditions, for all possible Hill's regions, we managed to classify the orbits into three types, namely bounded orbits around the primaries, escaping orbits, and orbits that lead to close encounters with the non-Newtonian primaries. Furthermore, the SALI chaos indicator has been used in order to further classify bounded orbits into three sub-categories: regular orbits, sticky orbits, and chaotic orbits. The transition from the Newtonian to pseudo-Newtonian regime has also been explored. More specifically, we determined how the Schwarzschild radius $r_S$, of the event horizon of the primaries, influences the dynamics of the system.

As far as we know, this is the first time that the nature of motion in a binary system of two black holes, is investigated in such a thorough and systematic manner, through the orbit classification using the two-dimensional color-coded diagrams. Obviously, our approach can be considered only as a first-order approximation to the general relativistic problem, taking into account that only some astrophysical features can be reproduced by the Paczy\'{n}ski-Wiita potential. Therefore we may claim that our paper adds considerable new information to the field of the non-Newtonian circular restricted three-body problem.

The following list contains the most important conclusions of our numerical analysis:
\begin{enumerate}
  \item Escaping motion completely dominates at both very low as well as very high levels of the total orbital energy. Furthermore, escaping motion is the only type of motion which is possible for all values of the energy.
  \item For intermediate values of the energy, we detected a very complicated mixture of all types of possible motion. In this regime, all the types of basins with the fractal basin boundaries exist.
  \item The vast majority of the initial conditions leading to bounded motion are regular orbits. In fact, there exist two main types of regular motion: (a) motion around one of the primary bodies and (b) motion around both primaries. Additional numerical computations revealed that regular motion mainly corresponds to simple loop orbits, while stability islands of secondary resonances are less common.
  \item The Schwarzschild radius $r_S$ was found to influence mostly close encounter orbits. In particular, the amount of the close encounter orbits increases rapidly, as the even horizon of the non-Newtonian primaries grows in size.
  \item The average escape and close encounter time of the orbits seems to be also connected with the Schwarzschild radius. In fact, for specific energy regions, the escape and close encounter times were found to reduce with increasing value of $r_S$.
\end{enumerate}

We hope that the current numerical results to be useful in the active field of relativistic astrophysics of binary systems.

\section*{Acknowledgments}

FLD and GAG gratefully acknowledge the financial support provided by COLCIENCIAS (Colombia), under Grants No. 8840 and 8863. The authors would like to express their warmest thanks to the anonymous referee for all the apt suggestions and comments which improved both the quality as well as the clarity of the paper.

\appendix

\section{Existence of circular orbit in the two-body problem}
\label{appex}

It is assumed that both primaries are free to move, each in a circle around the barycenter. The radius of each circle will be the distance from each primary to the barycenter, for the case of the Copenhagen problem (equal masses) $R/2$. Then, we calculate the gravitational force on each primary exerted by the other primary (derived from the Paczy\'{n}ski-Wiita potentials) and equate it to $m_i \omega^2 R/2$, where the gravitational force is calculated using the distance to the other primary $R$, i.e

\begin{align}
\frac{G m_1 m_2}{\left(R - r_S\right)^2} &= \frac{1}{2} m_1 \omega^2 R, \nonumber\\
\frac{G m_1 m_2}{\left(R - r_S\right)^2} &= \frac{1}{2} m_2 \omega^2 R.
\label{for}
\end{align}

Adding the previous expressions , we obtain
\begin{equation}
\omega^2 = \frac{R^2}{\left(R - r_S \right)^2}\frac{G\left(m_1 + m_2\right)}{R^3},
\label{omega1}
\end{equation}
which corresponds to Eq. (\ref{om1}).

Using a different approach, it is possible to show that the angular velocity is still valid for the combination of two Paczy\'{n}ski-Wiita potentials and that the
circular orbit for the binary indeed exists. To do so, we reduce the problem to that of the relative motion of the two primaries, here the barycenter moves at a constant velocity, and the linear momentum is conserved, reducing the degrees of freedom from six to three. A suitable choice of the inertial frame is at the barycenter, such that following the same procedure of the Newtonian problem, the Lagrangian reads as
\begin{equation}
\mathcal{L} = \frac{1}{2}\mu \left(\dot{r}^2 + r^2 \dot{\theta}^2 \right) + \frac{G \mu \mathcal{M}}{r - r_S},
\label{lag}
\end{equation}
with $\mu = m_1 m_2/\mathcal{M}$, and $\mathcal{M} = m_1 + m_2$. The respective equations of motion are given by
\begin{equation}
\ddot{r} - r \dot{\theta}^2 + \frac{G \mathcal{M}}{\left(r - r_S \right)^2} = 0,  \ \ \ \ L = \mu r^2 \dot{\theta},
\label{eqmot2}
\end{equation}
where $L$ stands for the total angular momentum, which is conserved. Now, circular orbits do exist for $r = const. = R$. Therefore from the previous equations we get
\begin{equation}
\frac{G \mathcal{M}}{\left(R - r_S\right)^2} = R \dot{\theta}^2.
\end{equation}
By setting $\dot{\theta} = \omega$ to the above equation we obtain
\begin{equation}
\omega^2 = \frac{R^2}{\left(R - r_S\right)^2} \frac{G \mathcal{M}}{R^3},
\label{omega2}
\end{equation}
which completely coincides with the result of Eq. (\ref{omega1}).

\bsp
\label{lastpage}

\end{document}